\documentclass[11pt, a4paper, logo]{googlecloud}

\pdftrailerid{redacted}

\makeatletter
\renewcommand\bibentry[1]{\nocite{#1}{\frenchspacing\@nameuse{BR@r@#1\@extra@b@citeb}}}
\makeatother

\RequirePackage{algorithm}
\RequirePackage{algorithmic}

\usepackage{kantlipsum, lipsum}
\usepackage{dsfont}

\usepackage[authoryear, sort&compress, round]{natbib}

\usepackage[utf8]{inputenc} 
\usepackage[T1]{fontenc}    
\usepackage[loose]{minitoc}
\usepackage{url}            
\usepackage{booktabs}       
\usepackage{nicefrac}       
\usepackage{microtype}      
\usepackage{amsmath}
\usepackage{graphicx}
\usepackage{multicol}
\usepackage{hyperref}       
\usepackage[nameinlink]{cleveref}
\usepackage{bbm}
\usepackage{multirow}
\usepackage{adjustbox}
\usepackage{listings}
\usepackage{subcaption}
\usepackage{soul}
\usepackage{float}
\usepackage{wrapfig}
\usepackage{blindtext}
\usepackage{tablefootnote}
\usepackage{amsfonts}
\usepackage[flushleft]{threeparttable}
\usepackage{bbding}
\usepackage{xcolor}
\usepackage{xspace}
\usepackage{bm}
\usepackage{arydshln}
\usepackage{enumitem}
\usepackage{setspace}
\usepackage{color}
\usepackage{longtable}

\usepackage[normalem]{ulem}
\usepackage{ulem}
\usepackage[nomargin,inline,marginclue,draft]{fixme}
\usepackage{balance}
\usepackage{verbatim}
\usepackage{diagbox}
\usepackage{changepage}
\usepackage{amssymb}
\usepackage{pifont}
\usepackage[section]{placeins}

\usepackage{array}   

\usepackage{makecell}
\usepackage{tabularray}

\usepackage{amsmath,amsfonts,bm}

\def\eqref#1{equation~\ref{#1}}

\def\1{\bm{1}}

\def\evd{{d}}

\DeclareMathAlphabet{\mathsfit}{\encodingdefault}{\sfdefault}{m}{sl}
\SetMathAlphabet{\mathsfit}{bold}{\encodingdefault}{\sfdefault}{bx}{n}

\def\sA{{\mathbb{A}}}

\def\sD{{\mathbb{D}}}

\def\sH{{\mathbb{H}}}

\def\sQ{{\mathbb{Q}}}

\usepackage{tcolorbox}

\definecolor{dkgreen}{rgb}{0,0.6,0}
\definecolor{gray}{rgb}{0.5,0.5,0.5}
\definecolor{mauve}{rgb}{0.58,0,0.82}

\usepackage{tcolorbox}

\definecolor{dkgreen}{rgb}{0,0.6,0}
\definecolor{gray}{rgb}{0.5,0.5,0.5}
\definecolor{mauve}{rgb}{0.58,0,0.82}

\definecolor{codegreen}{rgb}{0,0.6,0}
\definecolor{codegray}{rgb}{0.5,0.5,0.5}
\definecolor{codepurple}{rgb}{0.58,0,0.82}
\definecolor{backcolour}{rgb}{0.95,0.95,0.92}

\lstdefinestyle{mystyle}{
    backgroundcolor=\color{backcolour},   
    commentstyle=\color{codegreen},
    keywordstyle=\color{magenta},
    numberstyle=\tiny\color{codegray},
    stringstyle=\color{codepurple},
    basicstyle=\ttfamily\footnotesize,
    breakatwhitespace=false,         
    breaklines=true,                 
    captionpos=b,                    
    keepspaces=true,                 
    numbers=left,                    
    numbersep=5pt,                  
    showspaces=false,                
    showstringspaces=false,
    showtabs=false,                  
    tabsize=2
}
\title{LLM-based Multi-Agent Blackboard System for Information Discovery in Data Science}

\correspondingauthor{Alireza Salemi (asalemi@cs.umass.edu), Mihir Parmar (mihirparmar@google.com), Jinsung Yoon (jinsungyoon@google.com), and Hamid Palangi (hamidpalangi@google.com)}

\author[1]{Alireza Salemi$^{\dag}$}
\author[2]{Mihir Parmar}
\author[2]{Palash Goyal}
\author[2]{Yiwen Song}
\author[2]{Jinsung Yoon}
\author[1]{Hamed Zamani}
\author[2]{Tomas Pfister$^*$}
\author[2]{Hamid Palangi$^*$}

\affil[1]{University of Massachusetts Amherst}
\affil[2]{Google Cloud AI Research}

\begin{abstract}
  Advances in large language models (LLMs) have created new opportunities in data science, but their deployment is often limited by the challenge of finding relevant data in large data lakes. Existing methods struggle with this: both single- and multi-agent systems are quickly overwhelmed by large, heterogeneous files, and master–slave multi-agent systems rely on a rigid central controller that requires precise knowledge of each sub-agent’s capabilities, which is not possible in large-scale settings where the main agent lacks full observability over sub-agents’ knowledge and competencies. We propose a novel multi-agent paradigm inspired by the blackboard architecture for traditional AI models. In our framework, a central agent posts requests to a shared blackboard, and autonomous subordinate agents--either responsible for a partition of the data lake or retrieval from web--volunteer to respond based on their capabilities. This design improves scalability and flexibility by removing the need for a central coordinator to know each agent’s expertise or internal knowledge. We evaluate the approach on three benchmarks that require data discovery: KramaBench and modified versions of DSBench and DA-Code. Results show that the blackboard architecture substantially outperforms strong baselines, achieving 13\%–57\% relative improvements in end-to-end success and up to a 9\% relative gain in data discovery F1 over the best baseline. 
\end{abstract}

\begin{document}
\doparttoc 
\faketableofcontents 
\maketitle

\begingroup
    \renewcommand\thefootnote{\dag} 
    
    \footnotetext{Work done as a Student Researcher at Google.}
    \renewcommand\thefootnote{*}
    \footnotetext{Joint last authors.}
\endgroup

\setcounter{footnote}{0}

\section{Introduction}

\label{sec:introduction}

The recent developments in LLMs have introduced new paradigms for data science, enabling natural language-based approaches to data interpretation, transformation, and analysis \citep{jing2025dsbench, huang-etal-2024-da, hong-etal-2025-data, wang2025largelanguagemodelbaseddata}. Existing work, however, typically assumes an idealized setting in which relevant datasets are already curated and provided to the model---an assumption that diverges substantially from the practical challenges encountered in real-world data science \citep{lai2025kramabenchbenchmarkaisystems}. In practice, a substantial fraction of effort is devoted to locating the appropriate data within large and heterogeneous data lakes, often comprising thousands of loosely organized files---a process that constitutes a major bottleneck before any downstream analysis can be performed \citep{XU202132}. We argue that this stage of data discovery is both a critical and underexplored challenge for applying LLMs effectively. 

Previous work on data science tasks that require discovery\footnote{Examples are shown in Figure \ref{fig:search-example-1} and \ref{fig:search-example-2} in Appendix \ref{app:case-study}. They require computing/aggregating information from raw data within a data lake, where the specific source files are not pre-identified.} from a data lake has primarily relied on single-agent systems in which an LLM is given access to all candidate files within its context window and is then asked to solve the problem \citep{lai2025kramabenchbenchmarkaisystems}. This suffers from several limitations. First, it is not scalable: as the number of files grows, fitting them into the limited context of an LLM becomes infeasible. Second, the heterogeneity of files poses a challenge, as a single agent may struggle to effectively analyze and integrate diverse forms of information. Third, such systems lack robustness to noise, as the presence of many irrelevant files can overwhelm the model and degrade reasoning quality and precision. One may argue that Retrieval-Augmented Generation (RAG) \citep{10.5555/3495724.3496517, 10.1145/3626772.3657957} provides a solution by choosing a subset of files in the data lake; However, current retrievers are known to perform poorly on tabular and domain-specific data, which are pervasive in data science applications \citep{yu2025tableragretrievalaugmentedgeneration, ji2025targetbenchmarkingtableretrieval, huang-etal-2022-mixed, gu2025radarbenchmarkinglanguagemodels}.

\begin{figure}
    \centering
    \includegraphics[width=\textwidth]{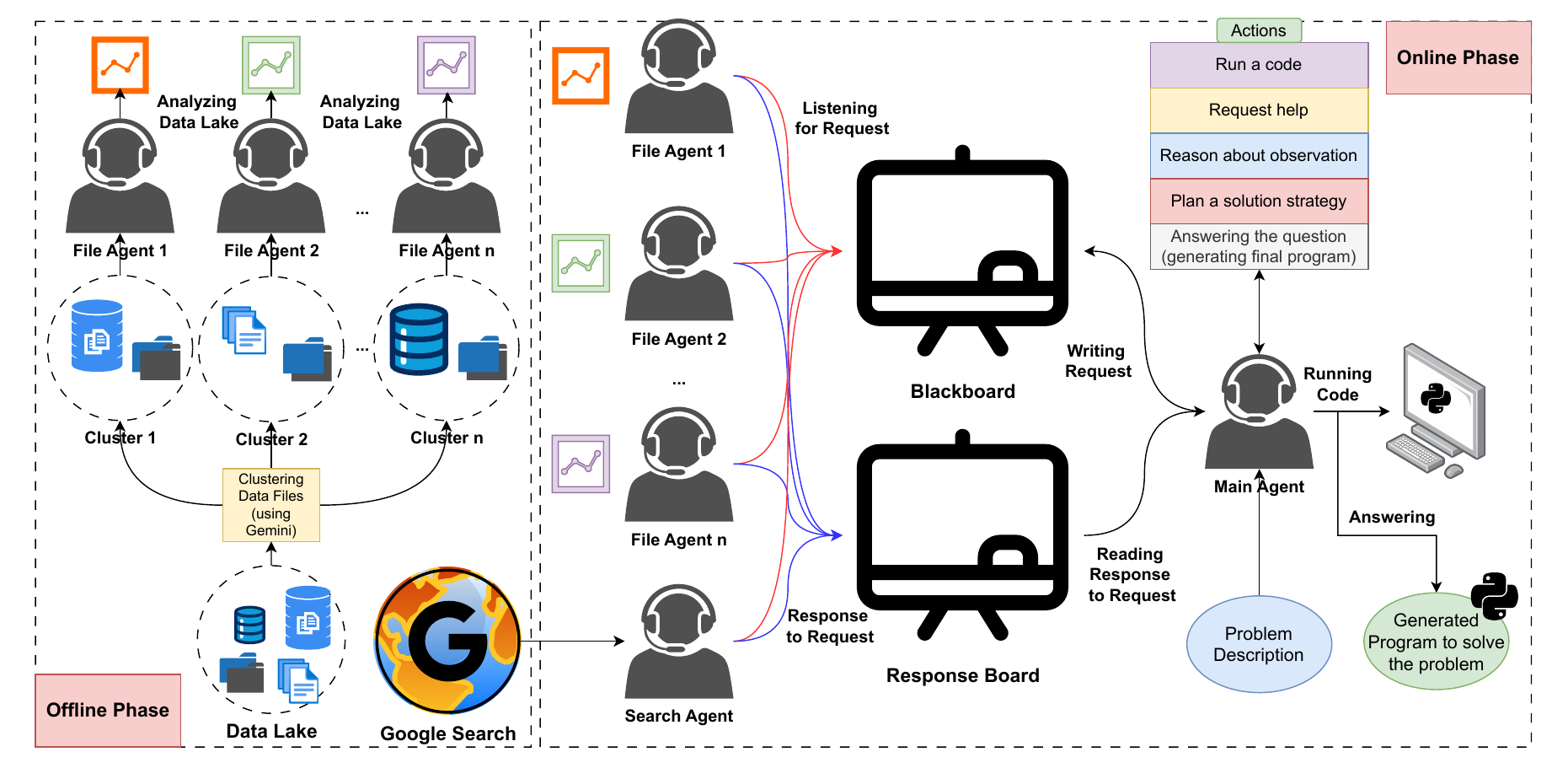}
    \caption{Overview of the blackboard multi-agent system for information discovery in data science. In this framework, the main agent does not assign tasks to subordinate agents. Instead, it posts requests to the blackboard, and subordinate agents autonomously decide whether to respond based on their expertise. The main agent then uses the responses to the request to solve the given task.}
    \label{fig:overview}
\end{figure}

An alternative approach explores multi-agent systems, which frequently adopt a master–slave paradigm \citep{li2025autokaggle, han2025llmmultiagentsystemschallenges, xu2025comprehensivesurveydeepresearch}. In this setting, a single controller (e.g., orchestration agent) assigns subtasks to a set of subordinate agents that then execute the specified actions. While conceptually straightforward, this architecture has several drawbacks. First, this master-slave paradigm limits the agents' autonomy: subordinate agents are forced to execute instructions from the coordinator even when they lack sufficient information or hold outdated or erroneous information. Second, the central controller must maintain an accurate model of each agents capabilities to assign tasks, an assumption that is often unrealistic when agents have only partial or evolving knowledge of the problem space. Finally, when multiple agents possess overlapping expertise, the controller faces an inherent assignment ambiguity, making task routing difficult. 

Inspired by the blackboard architecture that with substantial impact on traditional AI systems since the 1980s \citep{10.1145/356810.356816, blackboard-old-3}, we adopt a new communication paradigm for LLM multi-agent systems: A central agent remains responsible for solving the overall task, similar to the master-slave paradigm; However, rather than assigning subtasks to specific agents, the central agent posts a request on a shared \textit{blackboard} that describes the task or information needed, as shown in Figure~\ref{fig:overview}. Subordinate agents monitoring the blackboard can independently decide whether they possess the capability, knowledge, or interest to contribute to solving process. This design shifts decision-making from a single coordinator to a distributed model whose agents autonomously determine their participation, enabling more flexible collaborations. This differs from the conventional shared-memory paradigm in multi-agent systems. In shared-memory \citep{sagirova2025shared}, agents perform assigned tasks based on information in the shared memory, effectively being asked to execute them determined by a central coordinator. Conversely, \textit{in the blackboard architecture, there is no task assignment; instead, requests are broadcast on a blackboard, and each agent retains full autonomy whether to participate in the task or not.}

While the blackboard architecture can be applied broadly within multi-agent systems, its application to data science with data discovery is particularly compelling and underexplored. As shown in Figure~\ref{fig:overview}, the data lake can be partitioned into smaller clusters, e.g., based on similarity, homogeneity, or any criteria facilitating efficient handling, each assigned to a subordinate agent responsible for understanding and processing that subset. The main agent, which is tasked with solving the problem, posts requests on the blackboard specifying the data or general information required. Subordinate agents with the relevant knowledge or capability then autonomously volunteer to respond. This design ensures that each sub-agent manages only a subset of files or web-based information, enhancing scalability compared to approaches that require all data to be loaded into the main agent's prompt. Importantly, the main agent does not need prior knowledge of sub agents knowledge or capability, simplifying coordination and improving flexibility in large-scale data lakes. Here, the main agent's role is primarily to describe the information it requires and define tasks for sub agents, without directly managing or assigning it to them.

We conduct experiments on three datasets for data science requiring an explicit information discovery. KramaBench \citep{lai2025kramabenchbenchmarkaisystems} is a recently released benchmark and, to the best of our knowledge, the only publicly available dataset requires explicit data discovery in data science. We also repurpose DSBench \citep{jing2025dsbench} and DA-Code \citep{huang-etal-2024-da} by introducing a data discovery component, making them more challenging than their original formulations. Experimental results demonstrate that the blackboard architecture consistently outperforms strong baselines, including RAG, master-slave multi-agent system, and advanced data science baselines \citep{hong-etal-2025-data, wu2023autogenenablingnextgenllm}, achieving 13\% to 57\% relative improvement over the best performing baseline in end-to-end problem solving. This improvement is observed across both proprietary and open-source LLMs, highlighting the generalizability of the approach. Furthermore, our method also surpasses baselines in data discovery performance, yielding up to a 9\% relative gain in F1 score for correctly identifying relevant files from the data lake. These results underscore the effectiveness of the blackboard architecture as a communication paradigm for multi-agent systems in data science.

\section{Problem Formulation}

Let $\sD = \{\evd_{i}\}_{i=1}^{N}$ denote a data lake consisting of $N$ distinct data files, each containing information potentially completely or partially relevant to answering a data science question $q$ (some examples of these questions are shown in Figures~\ref{fig:search-example-1} and \ref{fig:search-example-2} in Appendix~\ref{app:case-study}). The objective of this work is to design a generative system $\pi_s$ that, given the query $q$ and the data lake $\sD$ as input, produces a program $p \sim \pi_s(q;\sD)$ in response. When executed (e.g., using a Python interpreter in this paper), this program $p$ retrieves, loads, and processes the appropriate data from the data lake $\sD$ and solve the given problem in the question $q$ to compute the answer. To evaluate the generated program $p$, we assume the existence of an evaluation function $\mu_{\text{generation}}$ that executes $p$ to produce an output $o_{p}$, compares $o_{p}$ with the ground-truth response $y_{q}$, and assigns a corresponding score. In addition, we assume a metric $\mu_{\text{retrieval}}$, given the program $p$ and the ground-truth files $\sD_{q}$, assigns a score reflecting the performance in discovering the correct data sources.

\section{LLM-based Multi-Agent Blackboard System}
\label{sec:method}

This section introduces an alternative paradigm for LLM-based multi-agent systems inspired by blackboard systems \citep{10.1145/356810.356816}, distinct from the widely used master–slave architecture. As outlined in \textsection \ref{sec:introduction}, blackboard multi-agent systems provide several advantages over the master-slave approach. Rather than directly assigning tasks to sub-agents, the main agent posts its requests (i.e., sub-tasks for which it requires assistance) on a blackboard as a broadcast channel accessible to all agents. Each helper agent independently evaluates whether it can respond to a request, considering its own capabilities, availability, and other factors. If an agent decides to contribute, it writes its response to the corresponding request, and the main agent then decides whether to use or ignore the provided information. \textit{This way, all agents in the system retain full autonomy over their actions, and no centralized controller forces them to execute a specific task.} While the blackboard paradigm is applicable to a wide range of tasks, we focus on data science tasks that require data discovery, where its characteristics are particularly advantageous, as discussed in \textsection \ref{sec:introduction}. The remainder of this section explains our method.

\paragraph{Overview:} 

An overview of our method is presented in Figure~\ref{fig:overview}. The system $\pi_{s}$ operates over the data lake $\sD$ by first partitioning $\sD$ into $C$ clusters of related files. Each cluster $\sD_i$ is assigned to a file agent $\pi_{f_i}$, which is responsible for handling, loading, processing, and retrieving information from the files within its cluster. In addition, a search agent $\pi_{s}$ is included to retrieve external information from the web that may be required to solve the problem. The overall system $\pi_{s}$ is composed of a main agent $\pi_{m}$, which is responsible for solving the query $q$, and a set of $C+1$ helper agents $\Pi_{\text{helper}} = \{\pi_{f_i}\}_{i=1}^{M} \cup \{\pi_{s}\}$ that provide specialized assistance. The query $q$ is presented to $\pi_{m}$, which iteratively selects an action $a \in \sA$ from the action space $\sA$, executes it, and observes the outcomes. Among its actions, the main agent may interact with a blackboard $\beta$, a shared communication medium where it can post a request $r$ without addressing a specific sub-agent. The helper agents $\Pi_{\text{helper}}$ continuously monitor the blackboard, determine whether they can address a request, and, if so, provide their outputs on the corresponding response board $\beta_{r}$. These responses are then collected and given to $\pi_{m}$, which incorporates them into its decision-making process.\footnote{Responses aren't written to board $\beta$ to prevent cross-influence of sub-agents; instead, they are written to a board $\beta_{r}$, enabling independent operation and exclusive access by the main agent.} The main agent is limited to at most $T$ sequential actions (including blackboard interactions) to solve the query $q$, ultimately producing a python program $p$ that answers $q$.

\paragraph{Clustering Data Lake:} 

There are different methods for partitioning the data lake into clusters; applying clustering algorithms over file representations, random partitioning, or other heuristic methods. For simplicity, by default, we do not utilize file content and instead rely solely on file names. Specifically, the file names are provided to an LLM---Gemini-2.5-Pro---which using the prompt shown in Figure~\ref{fig:clustering-prompt}, clusters the files into categories based only on their names. An example of this clustering is provided in Figure~\ref{fig:clustering-example} in Appendix~\ref{app:case-study}, where the model successfully groups related files together. For instance, it clusters all files originating from the National Interagency Fire Center into a category labeled ``NIFC Wildfire Statistics.'' The number of automatically derived clusters for each dataset is reported in Table~\ref{tab:stats} in Appendix~\ref{app:dataset}. To further demonstrate generalizability of our method to clustering approach (Appendix~\ref{app:results-analysis}), we conduct an experiment that uses file content embedding with E5-large \cite{wang2022text} with KMeans as an alternative clustering method to our default mode. The details of this experiment are described in Appendix~\ref{app:results-analysis}.


\subsection{Main Agent}
\label{sec:main-agent}

The primary role of the main agent is to solve the problem in collaboration with the helper agents. The main agent follows the ReAct framework \citep{yao2023react}, where at each step $t$, given the query $q$ and the history of actions and observations $\sH_{t-1}$, it first reasons about what is the best next action and selects an action from a predefined action space, executes the action, observes the outcome, and appends the resulting observation to update the history $\sH_{t}$.\footnote{In this work, model inputs, outputs, and observations are appended to the LLM prompt using its chat-based format.} The prompt used by the main agent is shown in Figure~\ref{fig:main-agent-blackboard-prompt} in Appendix~\ref{app:prompts}. The agent selects one of the following predefined actions in each step, executes them, and observe their outcomes:

\begin{itemize}[leftmargin=*]
    \item \textit{\textbf{Planning:}} The LLM decomposes the problem into smaller sub-problems and outlines a plan for addressing them. This action has no external effect on the environment but serves as an internal step to guide the LLM's problem-solving. In response, the system simply acknowledges the proposed plan and instructs the LLM to proceed.
    
    \item \textit{\textbf{Reasoning:}} The LLM focuses on a specific aspect of the problem and explains its reasoning, analysis, or interpretation of the available observations and steps taken so far in this process. Similar to the planning step, this action has no external effect on the environment but functions as an internal mechanism to guide the LLM's problem-solving process. In response, the system simply acknowledges the reasoning and prompts the LLM to continue.
    
    \item \textit{\textbf{Executing Code:}} The agent generates python code, which is executed using a python interpreter. If the code runs successfully, the resulting outputs are returned to the agent for observation; otherwise, the agent receives the corresponding error messages. This action enables the agent to explore the problem interactively, inspect data files, and experiment with them to gain a deeper understanding of their content and structure and how to process them.
    
    \item \textit{\textbf{Requesting Help:}} The agent formulates a request for assistance from the sub-agents, specifying, for example, the types of data files or information needed, or the resources required to apply a tool or solve a sub-problem. This request is posted on the blackboard $\beta$ for the helper agents. Once the sub-agents respond, if they respond, their responses on the response board $\beta_r$ are provided back to the main agent as the outcome of this action for observation and further use in its decision-making process.
    
    \item \textit{\textbf{Answering:}} The agent concludes the problem-solving process by generating a final program that produces the answer to the query. This action terminates the process, and the output of this step constitutes the final program $p$ generated by the system to address the problem.
\end{itemize}

\subsection{Helper Agents}
\label{sec:sub-agents}

In a data science, information discovery can typically be categorized into two tasks: (1) identifying the specific files that contain the data necessary to the problem, and (2) retrieving general knowledge about concepts relevant to the problem, such as domain-specific terms or details of particular algorithms and methods. To support these, our framework employs two types of helper agents:

\paragraph{File Agent:} 

Handling all files in a data lake with a single agent is infeasible: it involves a large number of files, many of which are lengthy and may exceed the agents context window; the files span diverse topics, which can confuse the agent and hinder effective reasoning; and accessing and processing all files simultaneously can be computationally expensive, leading to unnecessary overhead and slower problem-solving. Thus, in our framework each file agent is assigned responsibility for a subset of data files, as described earlier in the clustering procedure. In an offline phase, the file agent $\pi_{f_i}$ takes as input a subset of the data lake $\sD_{i}$ and operates through a two-step procedure. First, the agent selects a subset\footnote{When filenames indicate multiple files of the same data type over different time periods, the agent can infer the structure from a small representative sample rather than inspecting all files.} (or all) of the files to examine their content. The contents of them are presented to the agent for inspection (details of presentation are in Appendix~\ref{app:implementation}). Secondly, after observing the selected files, the agent analyzes them, learning how they are structured, what pre-processing or transformations may be required, and how they should be processed in general. An example of such an analysis is provided in Figure~\ref{fig:file-agent-analyze-example} in Appendix~\ref{app:case-study}. Then, in the online phase, the agent listens for requests from the main agent. Upon receiving a request, based on the analysis it did earlier, it determines whether it can contribute to answering it. If so, the agent generates a detailed plan specifying which files in $\sD_i$ are relevant, how they should be loaded in Python code, what libraries to use, the steps required for data processing, and samples from the data. The prompt used to guide the file agent is shown in Figure~\ref{fig:file-agent-prompt} in Appendix~\ref{app:prompts}. 

\paragraph{Search Agent:}

Certain data science problems require task-specific knowledge about algorithms or domain expertise that the LLM may not possess. Thus, we design a search agent that retrieves information from a search engine. This agent operates according to the prompt shown in Figure~\ref{fig:search-agent-prompt} in Appendix~\ref{app:prompts}. Given a request $r$ on the blackboard $\beta$, the agent first determines whether it is capable of addressing the request. It is specifically restricted to general web-based information retrieval and does not respond to requests involving access to local files. If the agent determines feasibility, it enters an iterative search process with a maximum of $T_{\text{search}} = 3$ steps. At each step $t$, the agent generates a set of queries $\sQ_{t}$, which are submitted to a search engine---Google Custom Search Engine\footnote{We use Google Custom Search Engine, configured to exclude all websites related to the datasets used to prevent data leakage.}---to retrieve $k=3$ webpage per query. The content of the webpages are then extracted using \textit{beautifulsoup} library\footnote{\url{https://pypi.org/project/beautifulsoup4/}} to be presented to the search agent. The extracted documents are then evaluated by the agent to determine whether they provide sufficient information to answer the request. If so, the agent generates a response to the request, which is posted to the response board $\beta_r$. If the information is insufficient, a new set of queries is generated to continue gathering relevant data from the web.

\section{Experiments}

\subsection{Experimental Setup}
\label{sec:exp-setup}

\paragraph{Benchmarks:}

To the best of our knowledge, KramaBench is the only public benchmark for data science problems that explicitly incorporates a data discovery phase, which we adopt in our evaluation. In addition, we repurpose two existing datasets, DSBench (data analysis task) \citep{jing2025dsbench} and DA-Code \citep{huang-etal-2024-da}, to include in this phase. Specifically, we manually filtered out all questions that do not require any data file for answering, as well as those that lack sufficient hints for data discovery.\footnote{E.g., questions that ask for a data science metric on a column without specifying the structure or content of the relevant file.} After filtering, we aggregated all remaining files across questions into a unified data lake, such that the model must perform discovery to identify relevant files at inference time. In this setup, only the question and the data lake are provided to the model, requiring it to identify the relevant files to answer the question, following the same protocol as KramaBench. Further details on this filtering process, along with dataset statistics in Table~\ref{tab:stats}, are provided in Appendix~\ref{app:dataset}.

\paragraph{Evaluation:}

We execute each generated program and compare its output against the ground-truth for the corresponding question. For each dataset, we adopt its standard evaluation protocol. For KramaBench, we use the official evaluation script in its repository.\footnote{\url{https://github.com/mitdbg/KramaBench}} For DA-Code, we likewise rely on the official evaluation script by authors.\footnote{\url{https://github.com/yiyihum/da-code}} For DSBench, we use the original evaluation method, in which an LLM serves as the judge. The generated programs output is compared against the reference using Gemini-2.5-Pro as the judge LLM, with the evaluation prompt shown in Figure \ref{fig:eval-ds-bench} in Appendix \ref{app:dataset}, producing a binary score.

\paragraph{Inference Setup:} We set the maximum actions of the main agent to $T = 10$. We use nucleus sampling \citep{Holtzman2020The} with a temperature of $0.1$ for more deterministic inference and default value for other hyperparameters. Proprietary models are accessed via Vertex AI, while open-source models are served by vLLM \cite{10.1145/3600006.3613165}. At each step, we cap the number of generated tokens at 8,192. We use Gemini-2.5-Pro and -Flash \citep{comanici2025gemini25pushingfrontier}, and Claude-4-Opus \citep{anthropic2025claude4} as the proprietary and Qwen3-Coder\footnote{\url{https://huggingface.co/Qwen/Qwen3-Coder-30B-A3B-Instruct}} with 30 billion parameters \citep{qwen3technicalreport} as the open-source LLMs. Experiments are conducted on 2 NVIDIA A100 (80GB VRAM) GPUs.

\paragraph{Baselines:} To evaluate our method against alternative approaches for solving data science problems involving data discovery, we compare it with the following baselines:
\begin{itemize}[leftmargin=*]
    \item \textit{\textbf{DS-GRU:}} We adopt the only existing baseline (to the best of our knowledge) for data discovery in data science problems, which appends all available files directly into the LLM prompt and attempts to solve the problem \citep{lai2025kramabenchbenchmarkaisystems}. This baseline uses a self-correction loop that retries when errors occur in generated codes. For details, we refer the reader to \citet{lai2025kramabenchbenchmarkaisystems}.
    
    \item \textit{\textbf{Retrieval-Augmented Generation (RAG):}} This retrieves the top 5 files\footnote{This number is chosen based on the average files needed to solve problems (1.6) and the context size of backbone LLMs used.} based on the file names and contents (presenting file content to the LLM is detailed in Appendix~\ref{app:implementation}) from the data lake using E5-large\footnote{\url{https://hf.co/intfloat/e5-large-v2}} \citep{wang2022text}, a 330M-parameter embedding model and use it to solve the problem. It then follows the same ReAct procedure as the main agent described in \textsection~\ref{sec:main-agent}, with two key modification: 1) the retrieved files contents are presented directly to the LLM's prompt and 2) the help-request action is replaced with a restricted action that permits requests only to the search agent. This design isolates the effect of replacing the file discovery mechanism with RAG, allowing a controlled study of its impact on performance. The prompt used is shown in Figure~\ref{fig:main-agent-rag-prompt} in Appendix~\ref{app:prompts}.
    
    \item \textit{\textbf{Master-Slave:}} This follows the same procedure as the main agent described in \textsection\ref{sec:main-agent}. The key difference is that, instead of requests on the blackboard, the agent directly invokes sub-agents (consisting of the search agent and the file agents as explained in \textsection\ref{sec:sub-agents}) by referencing their names and assign task to them. The prompt used for this baseline is shown in Figure~\ref{fig:main-agent-master-slave-prompt} in Appendix~\ref{app:prompts}.
\end{itemize}

\begin{table*}
    \centering
    \caption{Results on the KramaBench, DS-Bench, and DA-Code benchmarks. The best results for each LLM are highlighted in \textbf{bold}. The KramaBench categories are abbreviated: Arc. (Archaeology), Ast. (Astronomy), Bio. (Biomedical), Env. (Environment), Leg. (Legal), and Wild. (Wildfire).}
    \label{tab:main-result}
    \adjustbox{max width=\textwidth}{
    \begin{tabular}{ll c cccccc c c c c}
        \toprule
        & \multirow{2}{*}{\textbf{Method}}& \multirow{2}{*}{\textbf{LLM}} & \multicolumn{7}{c}{\multirow{1}{*}{\textbf{KramaBench}}} & \multirow{2}{*}{\textbf{\makecell{DS-\\Bench}}} & \multirow{2}{*}{\textbf{\makecell{DA-\\Code}}} & \multirow{2}{*}{\textbf{\makecell{Average \\ (macro)}}} \\
        \cmidrule{4-10}
        & & & Arc. & Ast. & Bio. & Env. & Leg. & Wild. & Average & & & \\
        \midrule
        
        (1) & {DS-GRU} & \multirow{4}{*}{\makecell{Qwen3-\\Coder}} & \textbf{0.00\%} & 1.80\% & 2.11\% & 1.15\% & 3.27\% & 13.54\% & 3.64\% & 0.00\% & 0.00\% & 1.21\% \\
        
        (2) & {RAG} &  & \textbf{0.00\%} & 3.16\% & 4.99\% & 0.54\% & 6.19\% & 16.93\% & 5.30\% & 6.32\% & 0.00\% & 3.87\% \\
        
        (3) & {Master-Slave} &  & \textbf{0.00\%} & 3.55\% & 3.39\% & \textbf{7.77\%} & \textbf{8.90\%} & 21.79\% & 7.56\% & 7.55\% & 0.00\% & 5.03\% \\
        
        \cmidrule{1-2} \cmidrule{4-13}
        
        (4) & {Blackboard} & & \textbf{0.00\%} & \textbf{7.69\%} & \textbf{7.85\%} & 4.47\% & 6.36\% & \textbf{23.97\%} & \textbf{8.39\%} & \textbf{14.22\%} & \textbf{1.11\%} & \textbf{7.90\%} \\
        
        \midrule
        \midrule
        
        (5) & {DS-GRU} & \multirow{4}{*}{\makecell{Gemini 2.5\\Flash}} & 0.00\% & \textbf{7.83\%} & 0.09\% & 10.93\% & 12.46\% & 13.34\% & 7.44\% & 5.53\% & 0.00\% & 4.32\% \\
        
        (6) & {RAG} & & \textbf{16.66\%} & 3.57\% & 13.98\% & \textbf{28.57\%} & 10.97\% & 33.67\% & 17.90\% & 22.92\% & \textbf{2.75\%} & 14.52\% \\
        
        (7) & {Master-Slave} & & \textbf{16.66\%} & 3.16\% & 13.98\% & 17.46\% & 21.75\% & 25.80\% & 16.46\% & 26.48\% & 0.55\% & 14.49\% \\
        
        \cmidrule{1-2} \cmidrule{4-13}
        
        (8) & {Blackboard} & & \textbf{16.66\%} & 3.57\% & \textbf{14.78\%} & 22.92\% & \textbf{27.09\%} & \textbf{41.04\%} & \textbf{21.01\%} & \textbf{28.06\%} & 0.55\% & \textbf{16.54\%} \\
        
        \midrule
        \midrule
        
        (9) & {DS-GRU} & \multirow{4}{*}{\makecell{Gemini 2.5\\Pro}} & 25.00\% & 6.69\% & 10.64\% & 27.47\% & 5.94\% & 39.36\% & 19.18\% & 3.95\% & 0.00\% & 7.71\% \\
        
        (10) & {RAG} & & \textbf{33.33\%} & 8.47\% & 32.53\% & 31.36\% & 25.55\% & 38.32\% & 28.26\% & 27.27\% & 0.00\% & 18.51\% \\
        
        (11) & {Master-Slave} & & \textbf{33.33\%} & 8.47\% & 24.74\% & 32.81\% & 34.64\% & 58.98\% & 32.16\% & 34.38\% & 5.49\% & 24.01\% \\
        
        \cmidrule{1-2} \cmidrule{4-13}
        
        (12) & {Blackboard} & & \textbf{33.33\%} & \textbf{17.95\%} & \textbf{36.83\%} & \textbf{39.31\%} & \textbf{34.92\%} & \textbf{62.88\%} & \textbf{37.53\%} & \textbf{38.73\%} & \textbf{9.34\%} & \textbf{28.53\%} \\
        
        \midrule
        \midrule
        
        (13) & {DS-GRU} & \multirow{4}{*}{\makecell{Claude 4\\Opus}} & 8.33\% & 1.38\% & 1.90\% & 8.14\% & 9.80\% & 23.14\% & 8.78\% & 3.55\% & 0.00\% & 4.11\% \\

        (14) & {RAG} &  & \textbf{33.33\%} & 11.52\% & 23.42\% & 31.61\% & 31.80\% & 45.80\% & 29.58\% & 35.57\% & 3.85\% & 23.00\% \\

        (15) & {Master-Slave} & & \textbf{33.33\%} & 8.69\% & 32.28\% & \textbf{39.16\%} & \textbf{44.08\%} & 48.35\% & 34.31\% & 45.84\% & 2.75\% & 27.63\% \\

        \cmidrule{1-2} \cmidrule{4-13}

        (16) & {Blackboard} &  & \textbf{33.33\%} & \textbf{18.69\%} & \textbf{45.31\%} & 34.35\% & 42.48\% & \textbf{50.06\%} & \textbf{37.37\%} & \textbf{49.80\%} & \textbf{7.14\%} & \textbf{31.43\%} \\
        \bottomrule
    \end{tabular}}
\end{table*}

\subsection{Empirical Findings}

\paragraph{Main Results:}

\begin{figure*}[!t]
    \centering
    \includegraphics[width=\textwidth]{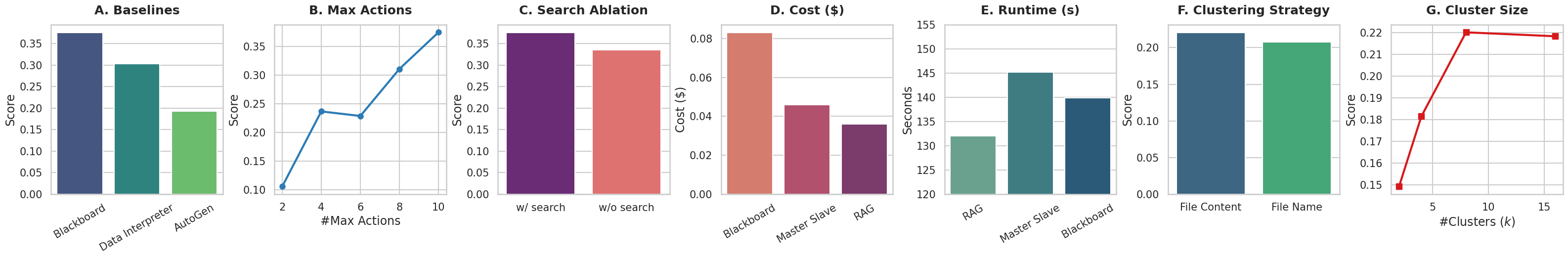}
    \vspace{-0.8cm}
    \caption{Performance and Efficiency Evaluation of the Blackboard System. \textbf{(A)} Performance vs. advanced baselines; \textbf{(B)} performance scalability with max actions; \textbf{(C)} impact of search component; \textbf{(D)} cost analysis per question; \textbf{(E)} runtime comparison showing efficiency gains; \textbf{(F)} content-based vs. filename clustering; and \textbf{(G)} optimization of cluster size $k$.}
    \vspace{-0.4cm}
    \label{fig:combined}
\end{figure*}

We conduct experiments on the datasets described in \textsection\ref{sec:exp-setup}. The results are presented in Table~\ref{tab:main-result}. These results demonstrate that our method, the Blackboard System, outperforms all baselines on average across all the datasets. Specifically, the Blackboard System surpasses the DS-GRU, RAG and Master-Slave approaches on all three datasets and achieves similar or higher performance in 4 out of 6 categories on KramaBench. Furthermore, we observe that the Blackboard System consistently outperforms the baselines regardless of the backbone LLM, highlighting its robustness and generalizability. We attribute this improvement to the design of the Blackboard System, where tasks are not explicitly assigned to helper agents; instead, each agent autonomously decides whether to participate based on its capabilities. This self-selection enhances both problem-solving efficiency and data discovery performance. We further compare the Blackboard system with two advanced methods for data analysis, Data Interpreter \citep{hong-etal-2025-data} and AutoGen \citep{wu2023autogenenablingnextgenllm}. Figure~\ref{fig:combined}(A) shows average performance on KramaBench, with per-task results in Table~\ref{tab:krama-gemini-pro} (Appendix~\ref{app:results-analysis}). The Blackboard system outperforms these baselines on five of six tasks and on average, indicating more effective adaptation and coordination across heterogeneous data science workloads.

\paragraph{File Discovery Performance:}

To analyze the effectiveness of different methods in data discovery, we report recall, precision, and F1-score for the file discovery task, i.e., identifying the correct files required to answer each question. The results of this experiment, using Gemini 2.5 Pro as the LLM, are presented in Table~\ref{tab:result-file-discovery}. The results indicate that the blackboard system achieves the highest recall, precision, and F1-score, both on average and across the three datasets. In particular, for KramaBench, the blackboard system attains the highest F1-score in 4 out of 6 domains. We attribute this improvement to the design of our method, where the main agent does not directly assign requests to specific file agents, as in the master–slave setup. Instead, each file agent independently decides whether it can contribute based on its data holdings, leading to more accurate file discovery.

\paragraph{Effect of Number of Main Agent's Actions on Performance:}

To evaluate the effect of the main agent’s maximum action limit, we vary it across ${2, 4, 6, 8, 10}$ and evaluate the blackboard system on KramaBench using Gemini 2.5 Pro as the backbone LLM. Results in Figure~\ref{fig:combined}(B) for average performance and Figure~\ref{fig:num-actions} in Appendix~\ref{app:results-analysis} for per-task results show that increasing the action budget improves performance. This improvement reflects the agent’s ability to explore the problem more thoroughly, consider alternative strategies, and generate higher-quality solutions.

\paragraph{Effect of Search Agent on Performance:}

We observed the LLM sometimes lacks the domain-specific knowledge or familiarity with specialized algorithms needed to fully solve problems. To address this, we include a search agent that retrieves relevant external information. Comparing the blackboard system with and without the search agent on KramaBench (Figure~\ref{fig:combined}(C) for average and Figure~\ref{fig:search-wo-search} in Appendix~\ref{app:results-analysis} for each task, using Gemini 2.5 Pro) shows that incorporating the search agent improves average performance. Analysis reveals that when the main agent encounters unfamiliar concepts, it requests information from the web, which the search agent retrieves, allowing problem-solving to continue effectively. Examples illustrating this behavior are in Figures~\ref{fig:search-example-1} and \ref{fig:search-example-2} in Appendix~\ref{app:case-study}, highlighting the search agent’s role when external domain knowledge is essential.

\begin{table}[]
    \centering
    \caption{File discovery performance reported using recall, precision, and f1-score. The results are obtained using Gemini 2.5 Pro as the backbone LLM. The best results are highlighted in \textbf{bold}.}
    \label{tab:result-file-discovery}
    \adjustbox{max width=\textwidth}{
    \begin{tabular}{ll l cccccc c c c c}
        \toprule
        &\multirow{2}{*}{\textbf{Method}} & \multirow{2}{*}{\textbf{Metric}} & \multicolumn{7}{c}{\textbf{KramaBench}} & \multirow{2}{*}{\textbf{DS-Bench}} & \multirow{2}{*}{\textbf{DA-Code}} & {\textbf{Average}} \\
        \cmidrule{4-10}
        & & & Archaeology & Astronomy & Biomedical & Environment & Legal & Wildfire & Average & & & (macro) \\
        \midrule
        \multirow{3}{*}{(1)} & \multirow{3}{*}{RAG } & recall & 0.875 & 0.125 & \textbf{0.666} & 0.3506 & 0.127 & 0.238 & 0.396 & 0.035 & 0.257 & 0.229 \\
        & & precision & \textbf{1.000} & 0.125 & 0.666 & 0.450 & 0.133 & 0.452 & 0.471 & 0.047 & 0.456 & 0.324  \\
        & & F1 & 0.916 & 0.125 & 0.629 & 0.332 & 0.105 & 0.301 & 0.401 & 0.034 & 0.307 & 0.247 \\
        \midrule
        \multirow{3}{*}{(2)} & \multirow{3}{*}{Master-Slave} & recall & \textbf{0.916} & 0.5138 & 0.648 & {0.382} & \textbf{0.444} & \textbf{0.567} & 0.578 & 0.323 & 0.546 & 0.482 \\
        & & precision & 0.930 & \textbf{0.750} & \textbf{0.722} & {0.500} & \textbf{0.494} & \textbf{0.642} & 0.673 & 0.503 & 0.767 & 0.647 \\
        & & F1 & 0.913 & 0.577 & \textbf{0.674} & 0.389 & \textbf{0.450} & \textbf{0.576} & 0.596 & 0.358 & 0.584 & 0.513 \\
        \midrule
        \multirow{3}{*}{(3)} & \multirow{3}{*}{Blackboard} & recall & \textbf{0.916} & \textbf{0.576} & 0.648 & \textbf{0.604} & 0.383 & 0.464 & \textbf{0.598} & \textbf{0.402} & \textbf{0.600} & \textbf{0.533} \\
        & & precision & \textbf{1.000} & 0.733 & \textbf{0.722} & \textbf{0.703} & 0.302 & 0.603 & \textbf{0.677} & \textbf{0.584} & \textbf{0.837} & \textbf{0.699} \\
        & & F1 & \textbf{0.944} & \textbf{0.618} & \textbf{0.674} & \textbf{0.588} & 0.304 & 0.495 & \textbf{0.603} & \textbf{0.438} & \textbf{0.643} & \textbf{0.561} \\
        \bottomrule
    \end{tabular}}
\end{table}

\paragraph{Scalability Comparison between Blackboard and Master–Slave Systems:}
We report relative (percentage) performance gains rather than absolute score differences to enable a fair comparison across datasets with varying difficulty and score ranges. Absolute improvements are not directly comparable across tasks because they depend on baseline accuracy and the attainable upper bound; for instance, a 5-point increase is minor when the baseline is already high but more meaningful for a challenging task with low initial performance. In contrast, relative gain measures the proportional improvement with respect to the baseline and therefore provides a normalized metric across heterogeneous tasks. Using this normalized measure, we analyze how the Blackboard architecture scales relative to the Master–Slave baseline as a function of data lake size (Figure~\ref{fig:performance-gain-datalake-size} in Appendix~\ref{app:results-analysis}). Each point in the figure corresponds to a task domain, and the regression line shows a positive correlation between data lake size and relative performance gain. Consistent with this trend, Blackboard achieves its largest gains on the two datasets with the largest data lakes, DA-Code (145 files) and Astronomy from KramaBench (1556 files), with relative improvements of 70\% and 211\%, respectively. These show that the Blackboard architecture scales more effectively in large data lakes, while the Master–Slave provides limited gains as the data lake grows.

\paragraph{Runtime and Cost Analysis:}

To characterize the efficiency–cost trade-off of the Blackboard system relative to the RAG and Master–Slave baselines, we randomly sample 50 questions from the KramaBench benchmark spanning all 6 different domains in the benchmark and measure both runtime and cost per question for each method. Unlike RAG and Master–Slave, which execute their component calls sequentially following the ReAct framework, the Blackboard architecture parallelizes sub-agent interactions: once the main agent posts a request to the shared blackboard, the corresponding sub-agents process it independently. As shown in Figure~\ref{fig:combined}(D, E) with more details in Figure~\ref{fig:runtime-cost} in Appendix~\ref{app:results-analysis}, the runtime of all three systems lies in a narrow range (132.0--145.2 seconds), indicating no substantial difference in latency. In terms of monetary cost, Blackboard is more expensive per question (approximately $2.3\times$ the cost of RAG and $1.8\times$ that of Master–Slave), due to increased token usage. However, this additional cost yields substantial performance gains—54.1\% over RAG and 18.8\% over Master–Slave—so that Blackboard achieves higher answer quality while maintaining comparable runtime, resulting in a favorable accuracy–cost trade-off.

\paragraph{Effect of Clustering Strategy and Number of Clusters on Performance:}

As described in \textsection\ref{sec:method}, we use Gemini 2.5 Pro to cluster files based on filenames, since providing full file contents is infeasible. To study content-based clustering, we instead encode file contents using the E5-Large embedding model and apply KMeans \citep{Lloyd1982LeastSQ}. Because KMeans requires a predefined number of clusters, we consider 2, 4, 8, and 16 clusters and restrict our analysis to datasets with at least 100 files (the Legal and Astronomy subsets of KramaBench and DA-Code). Results in Figure~\ref{fig:combined}(G) (average performance) and Figure~\ref{fig:cluster-num} in Appendix~\ref{app:results-analysis} (per dataset) show that increasing the number of clusters generally improves performance, as each sub-agent operates on a smaller, more coherent subset of files. Figure~\ref{fig:combined}(F) (average) and Figure~\ref{fig:file-vs-content} in Appendix~\ref{app:results-analysis} (per task) further compare filename-based clustering with Gemini to content-based clustering, showing that semantic content clustering outperforms the alternatives. The results show that our method supports embedding-based clustering and is not tied to a specific clustering approach.

\paragraph{Case Studies:}

To qualitatively analyze the blackboard system, we present several case studies:
\begin{itemize}[leftmargin=*, nosep]

\item \textbf{Requests on the blackboard:} An example request posted on the blackboard is shown in Figure~\ref{fig:request-example} in Appendix~\ref{app:case-study}. Given a data science question, the main agent formulates a request specifying likely column names, data formats, and interpretation guidance. In response, several helper agents (3 of 8 in this example) contribute. Although the relevant files are distributed across different clusters, each responding agent independently provides file locations, data-loading code, and descriptions of the data structure with suggested preprocessing steps. Together, these responses cover all ground-truth files needed to answer the question, illustrating how the blackboard mechanism enables effective information discovery and integration.

\item \textbf{Comparing Generated Program by Blackboard with Master-Slave System:} To further illustrate this difference, Figure~\ref{fig:program-example} in Appendix~\ref{app:case-study} compares programs generated by the Blackboard and Master--Slave systems. The Blackboard agent produces a better solution by correctly interpreting the prompt and selecting the appropriate data files: it identifies the patients' \texttt{Age} in \texttt{mmc1.xlsx} and the requested \texttt{APP-Z score} in \texttt{mmc7.xlsx}. In contrast, the Master--Slave agent misinterprets the request and uses a general protein abundance score (\texttt{APP\_log2\_abundance}) from \texttt{mmc2.xlsx}. This selection error leads to an incorrect result of \texttt{74}, whereas the Blackboard's accurate data discovery and reasoning yield the correct answer of \texttt{60}.

\end{itemize}

\section{Related Work}

Detailed related work is discussed in Appendix~\ref{app:related-work}.

\paragraph{LLMs for Data Science:} Specialized benchmarks evaluate LLMs in data science. DS-1000 \citep{10.5555/3618408.3619164}, ARCADE \citep{yin-etal-2023-natural}, DataSciBench \citep{zhang2025datascibenchllmagentbenchmark}, and DSEval \citep{zhang-etal-2024-benchmarking-data} focus on translating natural language instructions into correct implementations, unlike broader programming benchmarks such as SWE-Bench \citep{jimenez2024swebench}, ML-Bench \citep{tang2025mlbench}, and BigCodeBench \citep{zhuo2025bigcodebench}. Most assume relevant data files are pre-specified, though DSBench \citep{jing2025dsbench} and BLADE \citep{gu-etal-2024-blade} address multi-step reasoning, while ScienceAgentBench \citep{chen2025scienceagentbench} and BixBench \citep{mitchener2025bixbenchcomprehensivebenchmarkllmbased} focus on domain knowledge integration. These benchmarks largely overlook the challenge of discovering relevant data in large, heterogeneous repositories—a gap addressed by KramaBench \citep{lai2025kramabenchbenchmarkaisystems}, which explicitly evaluates data discovery. Building on this, we study how agents can autonomously identify and leverage correct data sources for end-to-end analysis. Applications of LLMs in data science have evolved from single-turn code generation to interactive, tool-augmented agents using models specialized for code, including GPT \citep{10.5555/3495724.3495883}, CodeGen \citep{rubavicius2025conversationalcodegenerationcase}, StarCoder \citep{li2023starcoder}, and Code Llama \citep{codellama}. While few-shot prompting \citep{10.5555/3495724.3495883} remains effective, state-of-the-art methods adopt agentic or multi-agentic frameworks combining reasoning with tool use. ReAct \citep{yao2023react} interleaves reasoning with actions, extended to execution environments \citep{chen2018executionguided}. Toolformer \citep{schick2023toolformer} and Gorilla \citep{patil2024gorilla} train LLMs to call APIs for tasks requiring specialized libraries. Self-correction frameworks like Self-Debug \citep{chen2024teaching} and Reflexion \citep{shinn2023reflexion} refine code using execution feedback. To improve reliability, many systems integrate RAG \citep{10.5555/3495724.3496517, salemi2025planandrefinediversecomprehensiveretrievalaugmented, 10.1145/3731120.3744584, 10.1145/3626772.3657733} to access documentation or code examples, reducing hallucinations and ensuring up-to-date library use. Multi-agent master–slave frameworks, such as AutoKaggle \citep{li2025autokaggle}, further demonstrate promising results in this domain.

\paragraph{Blackboard Systems:}

The blackboard system is a architectural model from classical AI, developed for problems that require incremental and opportunistic reasoning. It was implemented in the Hearsay-II speech understanding \citep{10.1145/356810.356816} and is characterized by three components: (1) a global, hierarchical data structure (the blackboard) that maintains the current state of the solution; (2) independent specialist modules, known as knowledge sources, which monitor the blackboard and contribute partial solutions; and (3) a control mechanism that opportunistically determines which knowledge source to activate next \citep{Nii_1986}. Following successful applications in domains such as sonar interpretation with the HASP/SIAP system \citep{Nii_Feigenbaum_Anton_1982}, the architecture evolved to incorporate more sophisticated control strategies. Inspired by this, we adapt the blackboard architecture for LLM-based multi-agent communication. Instead of a central controller assigning tasks, agents operate autonomously and respond to requests posted on the blackboard when they can contribute without being forced to participate. The main agent then integrates the information provided by the sub-agents to solve the problem.
\section{Conclusions \& Future Work}

We addressed the critical challenge of data discovery in large, heterogeneous data lakes, a key bottleneck for applying LLMs in data science. We introduced a novel multi-agent communication paradigm based on the blackboard architecture, which replaces rigid centralized task assignment with a flexible, decentralized model of agent collaboration. Extensive experiments on three data science benchmarks demonstrate that our framework consistently outperforms strong baselines, including RAG and the master–slave paradigm, achieving up to 57\% relative improvement in end-to-end task success and a 9\% relative gain in data discovery accuracy. These results highlight the importance of communication architecture in multi-agent systems and establish the blackboard paradigm as a scalable, flexible, and effective solution for complex data science workflows. Future work could extend the blackboard architecture beyond data science, as the proposed approach is general and applicable to a wide range of multi-agent systems. Another promising direction is to investigate more adaptive strategies for data partitioning among agents, enabling the system to better handle dynamic and evolving data environments. Ultimately, our findings point toward a broader path for developing more capable, scalable, and autonomous multi-agent AI systems for real-world data analysis applications.

\newpage

\bibliographystyle{abbrvnat}
\nobibliography*
\bibliography{ref}

\newpage
\appendix
\renewcommand{\partname}{}
\renewcommand{\thepart}{}
\addcontentsline{toc}{section}{Appendix} 
\part{Appendix}
\parttoc
\newpage
\let\clearpage\relax
\section{Datasets and Preprocessing}
\label{app:dataset}

To the best of our knowledge, KramaBench \citep{lai2025kramabenchbenchmarkaisystems} is the only publicly available dataset for data science problems that explicitly require data discovery to answer the questions. We adopt this dataset as one of our evaluation benchmarks in this paper. To further investigate this problem, we repurpose two widely used datasets for data science tasks, DSBench \citep{jing2025dsbench} and DA-Code \citep{huang-etal-2024-da}, which were not originally designed to include a data discovery phase. In their original form, each question in these datasets is paired with the specific data files required to answer it. To adapt them to our setting, we remove this direct mapping: the model is provided only with the question, while all files from the dataset are aggregated into a single data lake. The model must therefore first identify the relevant files within the data lake and then use them to solve the question.

\paragraph{Filtering:} we observed that not all questions in these datasets are suitable for the data discovery setting. For instance, some questions provide no hints about the characteristics of the files needed to answer them, while others simply ask for computing a statistic on a column without specifying sufficient information to identify the relevant file. To address this issue, we manually filter out such questions and retain only those that include adequate cues for discovering the appropriate files. We exclude questions that request performing a specific operation on a particular column of a data file when the column’s meaning or semantics are insufficiently described. In such cases, it would be infeasible to accurately identify the target column within the data lake, given that multiple files may contain columns with the same name. Furthermore, questions that focus solely on the operation itself—assuming access to only a single file—are also excluded, as they lack sufficient contextual information for meaningful retrieval or reasoning about the data file that needs to be discovered from the data lake. Finally, since our goal is to study information discovery in data science, we also exclude questions that can be answered without accessing any data files, as these are general data science questions not relevant to any data files. After this filtering process, the resulting dataset statistics are reported in Table~\ref{tab:stats}. Finally, we provide the example IDs corresponding to the data instances retained from each dataset (DSBench and DA-Code) in Appendix~\ref{app:dataset-ids}.

\begin{table}[h!]
    \centering
    \caption{Statistics of the datasets used in our evaluation setup.}
    \label{tab:stats}
    \begin{tabular}{l|ccc}
        \toprule
        \textbf{Dataset} & \textbf{\#Tasks} & \textbf{Size of data lake} & \textbf{\#Clusters created by Gemini 2.5 Pro} \\
        \midrule
        KramaBench & 104 & 1746\tablefootnote{Note that, in line with the original benchmark design, we construct a separate data lake for each subtask. However, the reported number of files corresponds to the total number of files aggregated across all subtasks in the benchmark.} & 27\tablefootnote{Note that, in line with the original benchmark design, we construct a separate data lake for each subtask. However, the reported number of clusters corresponds to the total number of clusters aggregated across all subtasks in the benchmark.} \\
        \quad\quad - Archeology & 12 & 5 & 3 \\
        \quad\quad - Astronomy & 12 & 1556 & 8 \\
        \quad\quad - Biomedical & 9 & 7 & 2 \\
        \quad\quad - Environment & 20 & 37 & 4 \\
        \quad\quad - Legal & 30 & 136 & 4 \\
        \quad\quad - Wildfire & 21 & 23 & 6 \\
        \midrule
        DSBench & 253 & 48 & 12 \\
        \midrule
        DA-Code & 91 & 145 & 26 \\
        \bottomrule
    \end{tabular}
\end{table}

\paragraph{Evaluation:}

\begin{figure}[h!]
    \centering
    \includegraphics{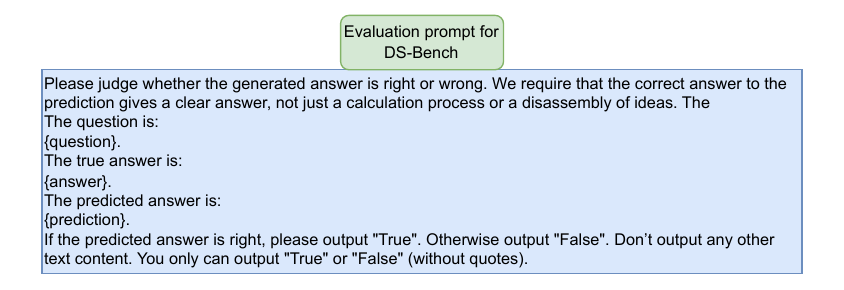}
    \caption{Evaluation prompt used for DSBench dataset using LLM as the judge.}
    \label{fig:eval-ds-bench}
\end{figure}

To evaluate the programs generated by the system, we execute each program and assess its final output against the reference answer for the given question. For each dataset, we adopt its original evaluation methodology. Specifically, for KramaBench, we use the official evaluation script provided in their repository.\footnote{Available at: \url{https://github.com/mitdbg/KramaBench}} For DA-Code, we similarly rely on the official evaluation script released in their repository.\footnote{Available at: \url{https://github.com/yiyihum/da-code}} For DSBench, we follow the original evaluation protocol that uses LLM-based judging: the generated programs output is compared to the reference answer using Gemini 2.5 Pro as the judge LLM, with the prompt shown in Figure~\ref{fig:eval-ds-bench}.

\newpage
\section{Detailed Related Work}
\label{app:related-work}

\paragraph{LLMs for Data Science:} Specialized benchmarks have emerged to evaluate LLMs in data science. DS-1000 \citep{10.5555/3618408.3619164}, ARCADE \citep{yin-etal-2023-natural}, DataSciBench \citep{zhang2025datascibenchllmagentbenchmark}, and DSEval \citep{zhang-etal-2024-benchmarking-data} assess the translation of natural language instructions into correct implementations, distinguishing them from broader programming benchmarks such as SWE-Bench \citep{jimenez2024swebench}, ML-Bench \citep{tang2025mlbench}, and BigCodeBench \citep{zhuo2025bigcodebench}. While most assume that the relevant data files are pre-specified, recent efforts address multi-step reasoning: DSBench \citep{jing2025dsbench} and BLADE \citep{gu-etal-2024-blade} evaluate implementation planning, and ScienceAgentBench \citep{chen2025scienceagentbench} and BixBench \citep{mitchener2025bixbenchcomprehensivebenchmarkllmbased} focus on integrating domain knowledge. These benchmarks, however, still overlook the practical challenge of discovering relevant data within large, heterogeneous repositories—a gap addressed by KramaBench \citep{lai2025kramabenchbenchmarkaisystems}, which explicitly evaluates data discovery. Building on this, we study how agents can autonomously identify and leverage the correct data sources for end-to-end analysis.

Applications of LLMs in data science have evolved from single-turn code generation to interactive, tool-augmented agents that exploit models specialized for code, including GPT \citep{10.5555/3495724.3495883}, CodeGen \citep{rubavicius2025conversationalcodegenerationcase}, StarCoder \citep{li2023starcoder}, and Code Llama \citep{codellama}. While few-shot prompting \citep{10.5555/3495724.3495883} remains effective, state-of-the-art approaches adopt agentic or multi-agentic frameworks that combine iterative reasoning with external tool use. ReAct \citep{yao2023react} pioneered the interleaving of reasoning and action, later extended to execution environments \citep{chen2018executionguided}. Toolformer \citep{schick2023toolformer} and Gorilla \citep{patil2024gorilla} explicitly train LLMs to call APIs, a capability critical for tasks relying on specialized libraries. Self-correction is a another key feature: frameworks like Self-Debug \citep{chen2024teaching} and Reflexion \citep{shinn2023reflexion} refine generated code using execution feedback. To further enhance reliability, many systems integrate RAG \citep{10.5555/3495724.3496517, salemi2025planandrefinediversecomprehensiveretrievalaugmented, 10.1145/3731120.3744584, 10.1145/3626772.3657733} to retrieve documentation or code examples, reducing hallucinations and ensuring up-to-date library use. Additionally, multi-agent master-slave frameworks, such as AutoKaggle \citep{li2025autokaggle}, have demonstrated promising results in addressing these challenges.

\paragraph{Blackboard Systems:}

The blackboard system is a seminal architectural model from classical AI, developed for complex problems that require incremental and opportunistic reasoning. It was implemented in the Hearsay-II speech understanding \citep{10.1145/356810.356816} and is characterized by three components: (1) a global, hierarchical data structure (the blackboard) that maintains the current state of the solution; (2) independent specialist modules, known as knowledge sources, which monitor the blackboard and contribute partial solutions; and (3) a control mechanism that opportunistically determines which knowledge source to activate next \citep{Nii_1986}. Following successful applications in domains such as sonar interpretation with the HASP/SIAP system \citep{Nii_Feigenbaum_Anton_1982}, the architecture evolved to incorporate more sophisticated control strategies. Inspired by this paradigm, we adapt the blackboard architecture for multi-agent communication: rather than a central controller assigning tasks, all agents operate autonomously, responding to requests posted on the blackboard. A central main agent then leverages the information contributed by sub-agents to solve the problem.

In parallel to our work, \citet{han2025exploringadvancedllmmultiagent} also study blackboard architectures, primarily for mathematical and reasoning tasks that do not require data discovery. In their design, the main agent is still responsible for assigning tasks to other agents by determining which agent should act next. The blackboard in their framework is mainly used as a shared memory for information exchange, while the main agent retains most of the decision-making responsibility. In contrast, in our framework, all agents operate autonomously, and no explicit task assignment is required. As discussed in the paper, this architecture is not suitable for our setting, since it is not possible to include all relevant information in the LLM’s prompt.

\newpage
\section{Agents' Prompts}
\label{app:prompts}

\begin{figure}[h!]
    \centering
    \includegraphics[width=0.85\textwidth]{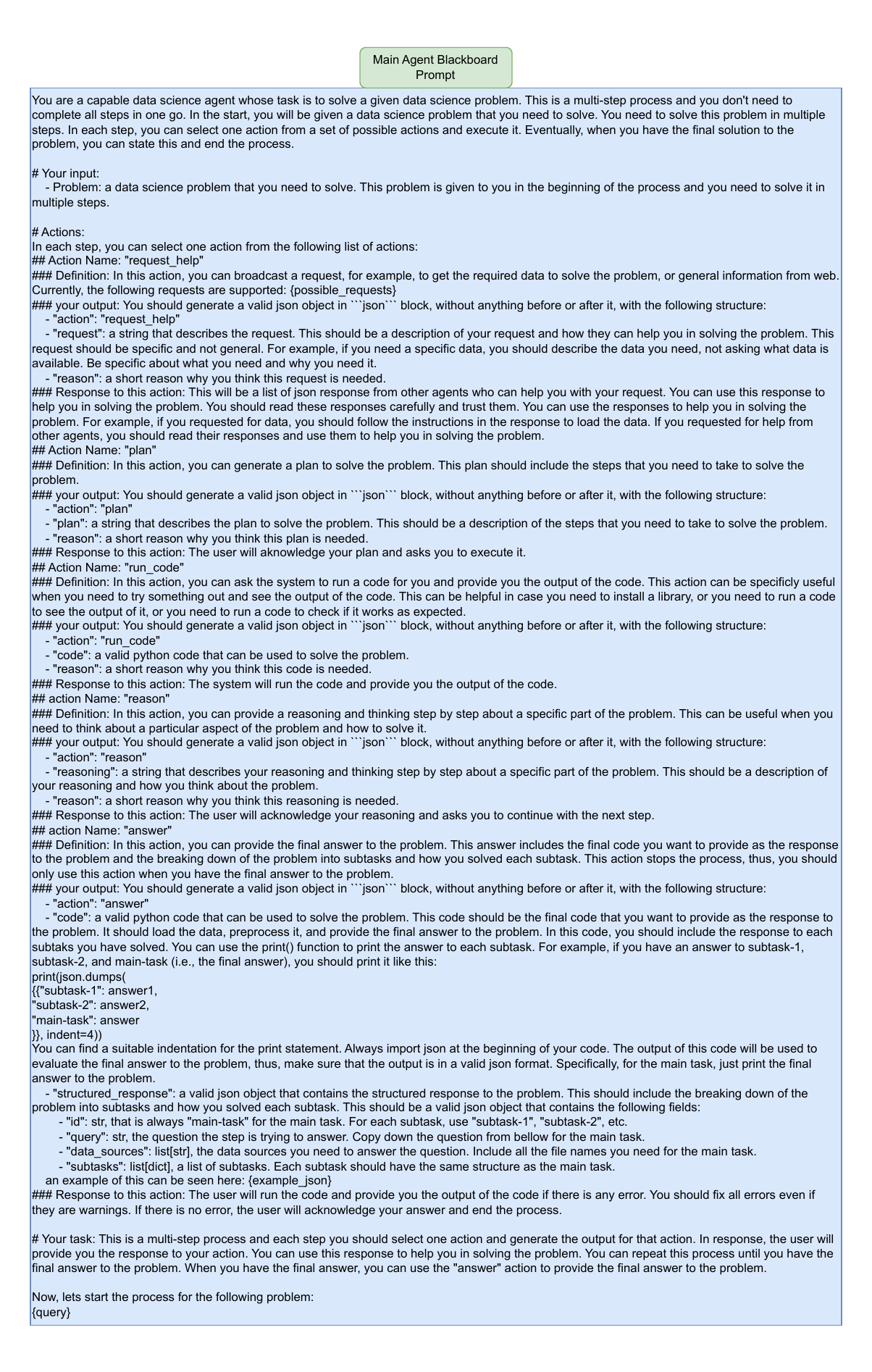}
    \vspace{-0.8cm}
    \caption{Prompt used by the main agent for the blackboard system.}
    \label{fig:main-agent-blackboard-prompt}
\end{figure}

\begin{figure}[h!]
    \centering
    \includegraphics[width=\textwidth]{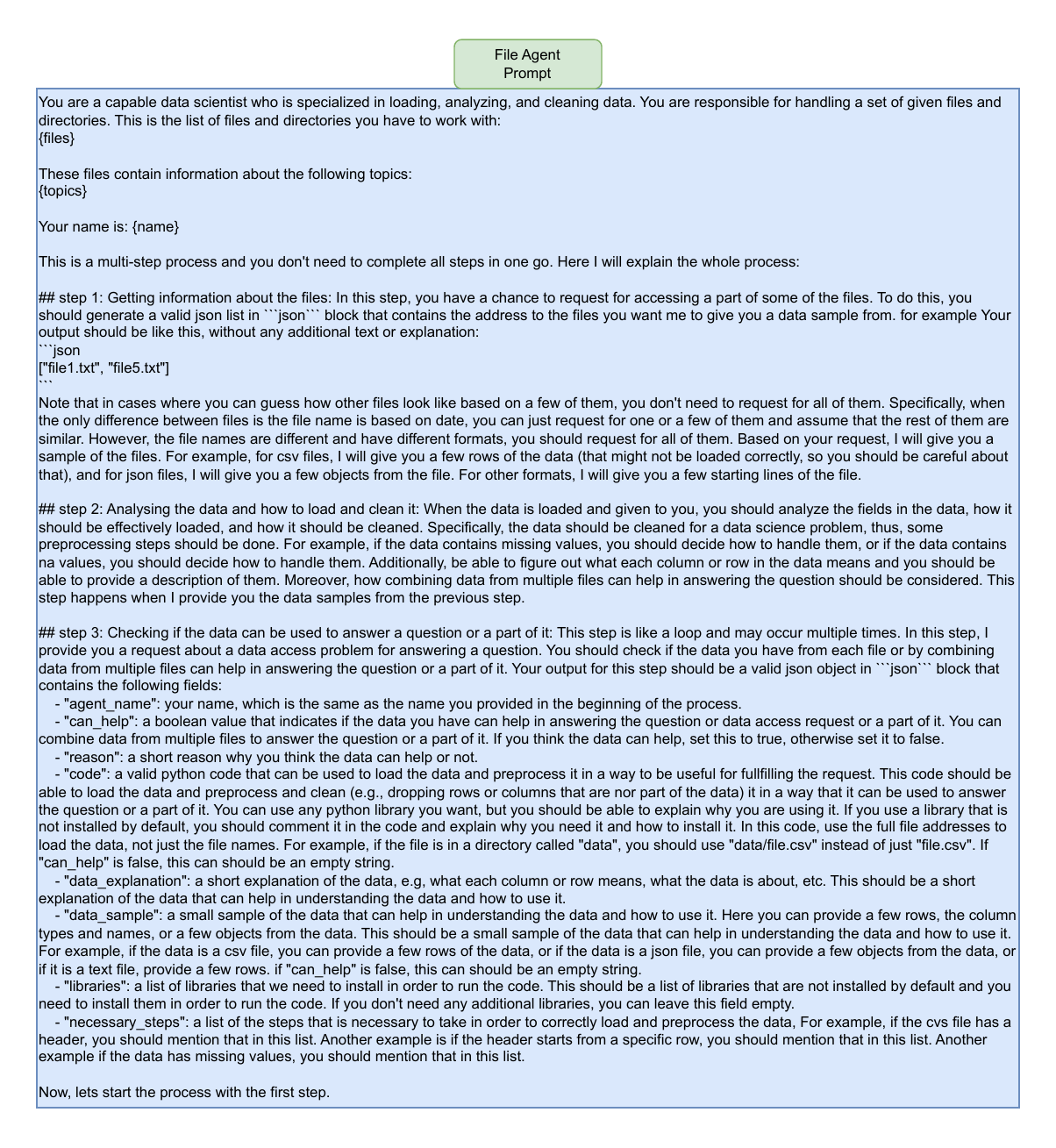}
    \vspace{-1.2cm}
    \caption{Prompt used by the file agent for the both master-slave and blackboard system.}
    \label{fig:file-agent-prompt}
\end{figure}

\begin{figure}[h!]
    \centering
    \includegraphics[width=\textwidth]{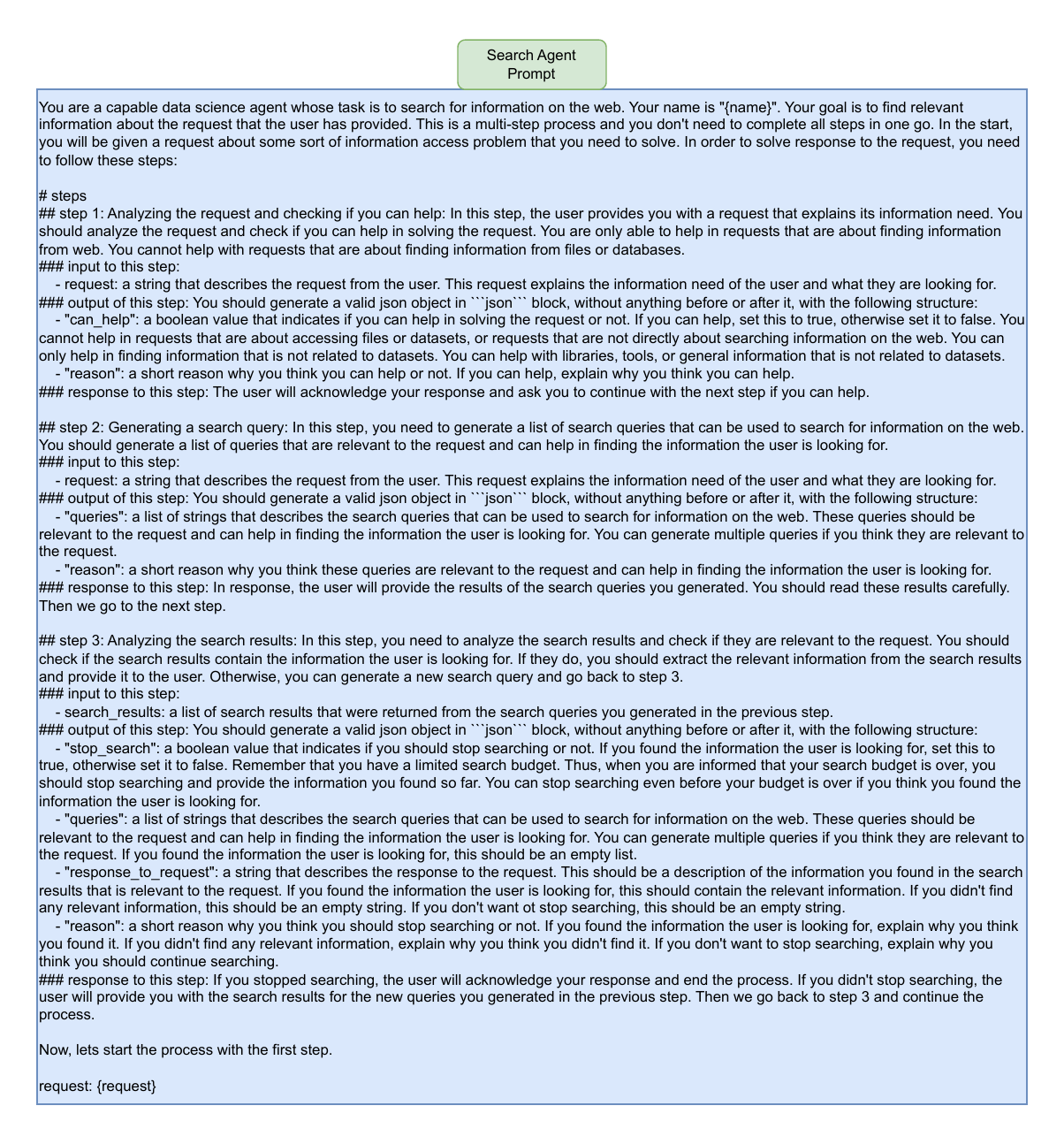}
    \vspace{-1.2cm}
    \caption{Prompt used by the search agent for the, master-slave, RAG, and blackboard system.}
    \label{fig:search-agent-prompt}
\end{figure}

\begin{figure}[h!]
    \centering
    \includegraphics[width=0.82\textwidth]{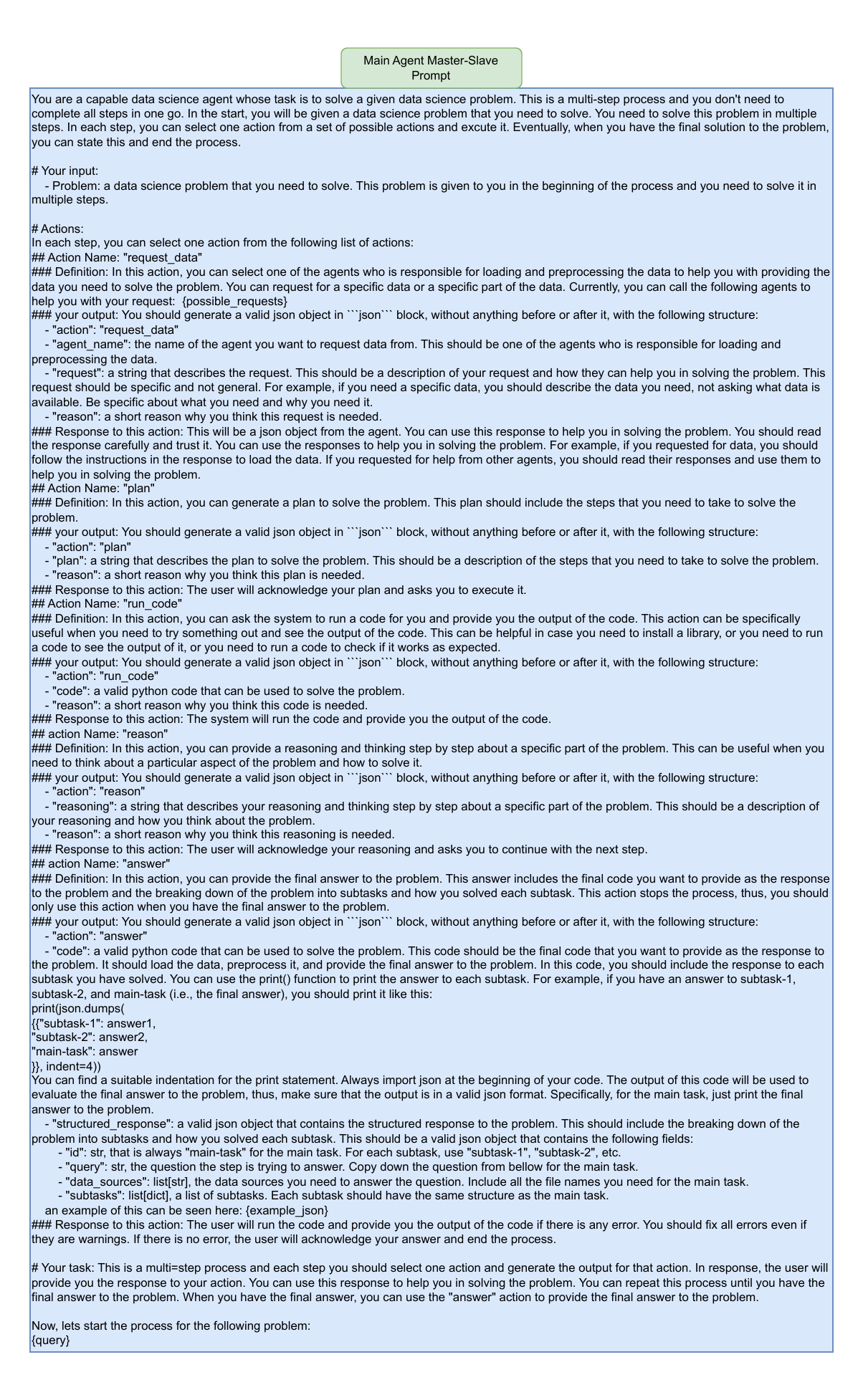}
    \vspace{-1cm}
    \caption{Prompt used by the main agent for the master-slave system.}
    \label{fig:main-agent-master-slave-prompt}
\end{figure}

\begin{figure}[h!]
    \centering
    \includegraphics[width=0.82\textwidth]{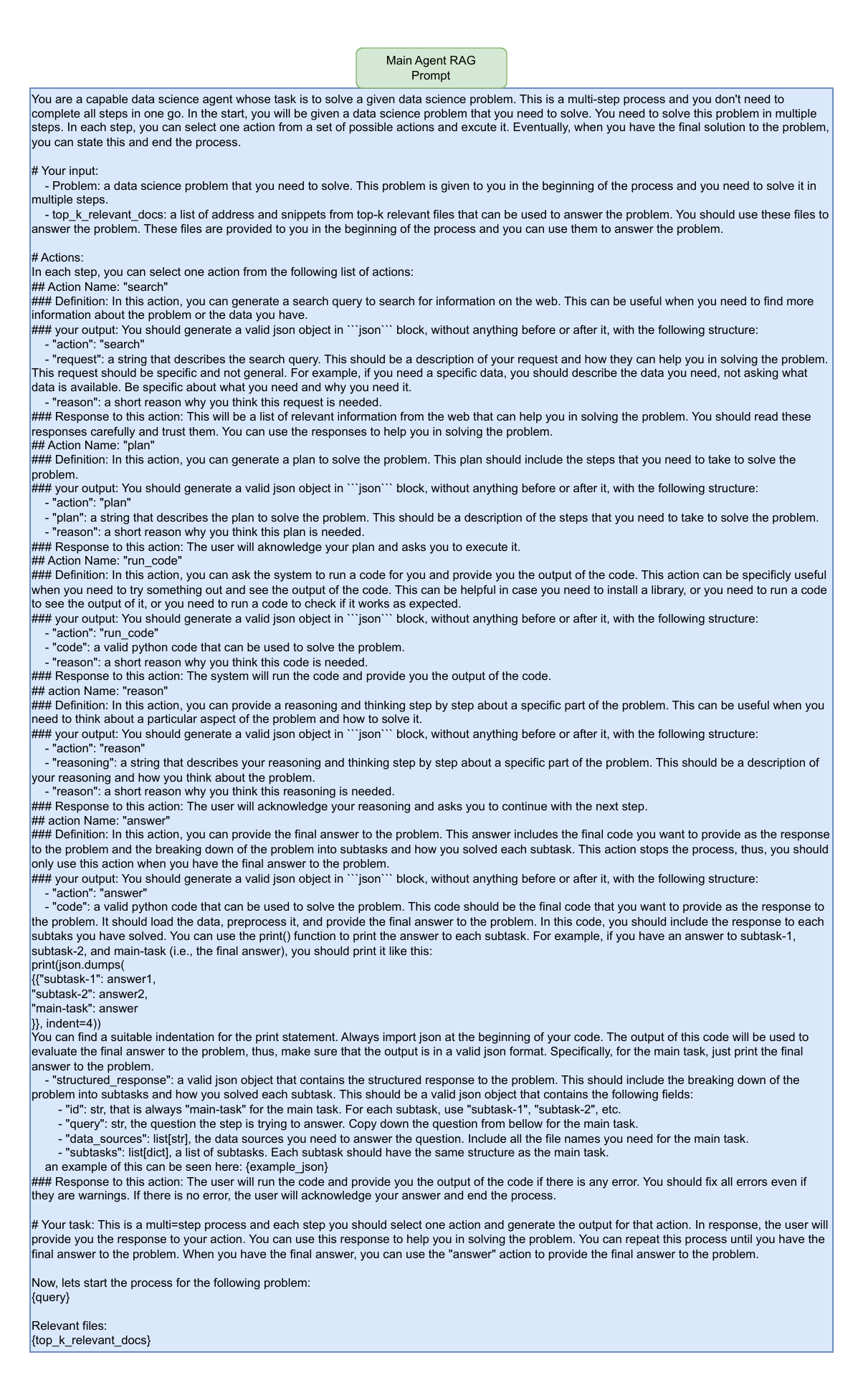}
    \vspace{-1cm}
    \caption{Prompt used by the main agent for the RAG system.}
    \label{fig:main-agent-rag-prompt}
\end{figure}

\begin{figure}
    \centering
    \includegraphics[width=0.9\textwidth]{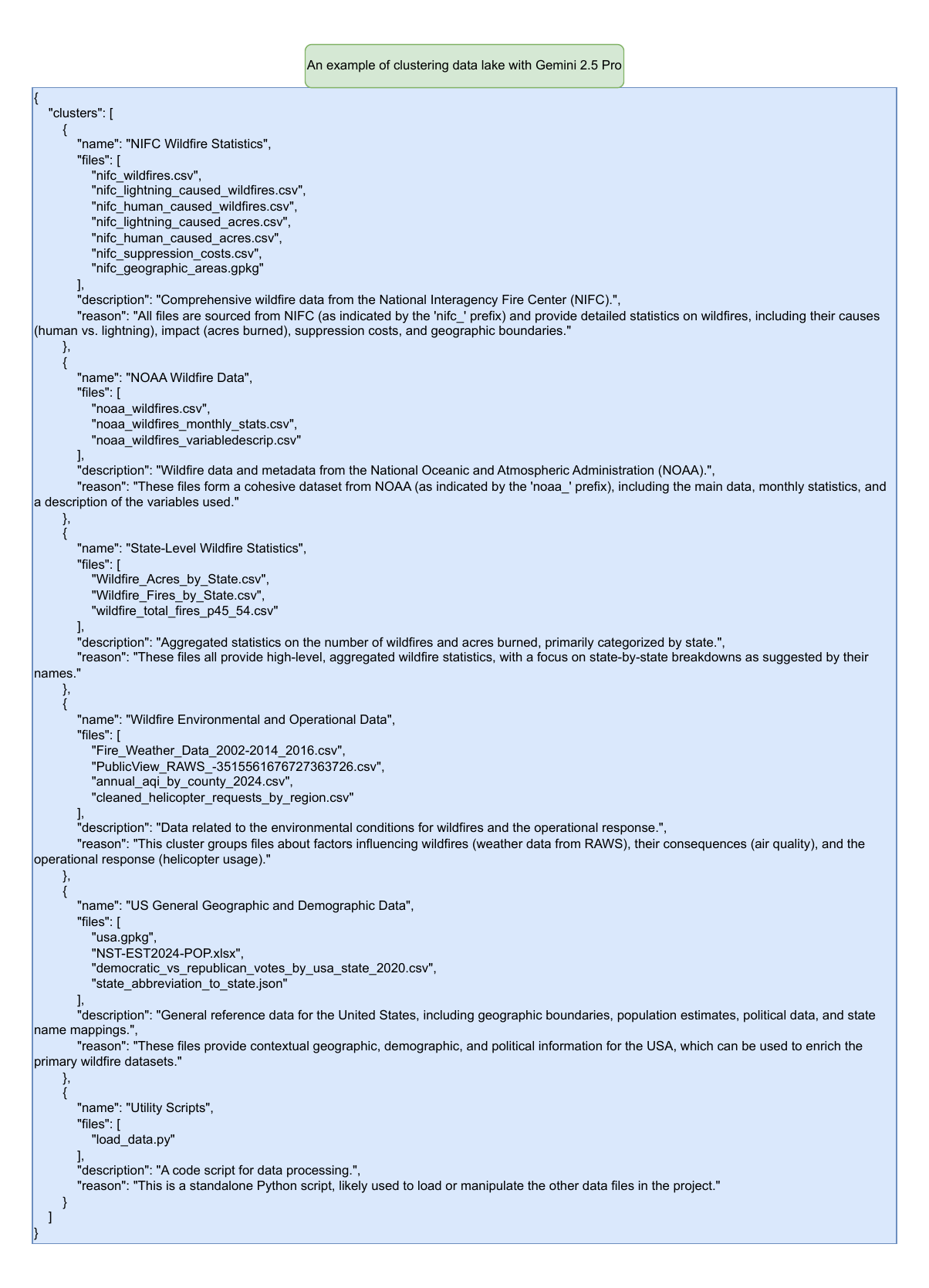}
    \vspace{-0.8cm}
    \caption{An example of clustering data lake into partitions using Gemini 2.5 Pro.}
    \label{fig:clustering-example}
\end{figure}

This section presents the prompts used by the agents and baselines in this paper. Figure~\ref{fig:clustering-example} shows the prompt for clustering the data lake into multiple partitions based on file names. Figure~\ref{fig:main-agent-blackboard-prompt} presents the prompt used by the main agent in the blackboard system. Figure~\ref{fig:file-agent-prompt} shows the prompt for the file agents. Figure~\ref{fig:search-agent-prompt} displays the prompt used by the search agent. Figure~\ref{fig:main-agent-master-slave-prompt} presents the prompt for the main agent in the master–slave system, and Figure~\ref{fig:main-agent-rag-prompt} shows the prompt used by the RAG agent.

\begin{figure}[h!]
    \centering
    \includegraphics[width=\textwidth]{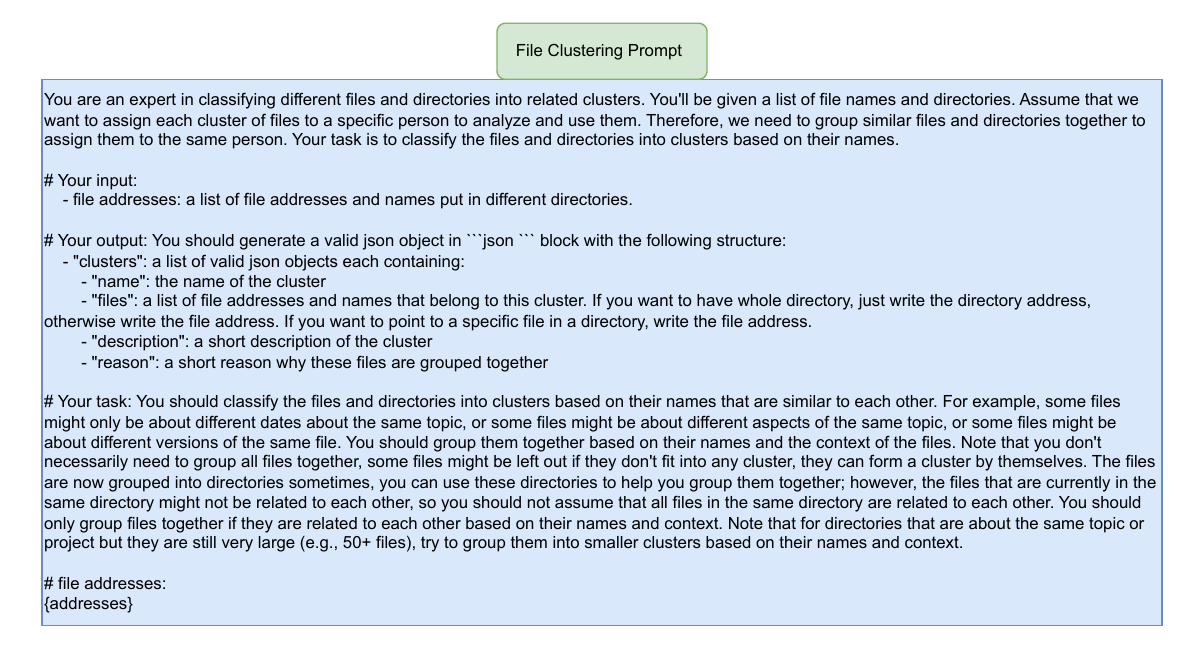}
    \vspace{-0.8cm}
    \caption{Prompt used by for clustering the files in data lakes into partitions.}
    \label{fig:clustering-prompt}
\end{figure}

\newpage

\newpage

\newpage

\newpage

\newpage

\newpage

\FloatBarrier
\section{Implementation Details}
\label{app:implementation}

\paragraph{Presenting Files to File Agents:} A file agent may request a file by name, in which case it is shown a subset of the files contents. For this case, we employ a controlled procedure for loading and presenting the data to the agent, as described below:
\begin{itemize}[leftmargin=*]

\item Files with \textit{.csv} format: In this case, we use the \textit{pandas}\footnote{Available at: \url{https://pandas.pydata.org/}} library to load the CSV files, presenting the column names, their data types, and the top 20 rows of the table to the agent.

\item Files with \textit{.gpkg} format: which provides a pandas-like interface for geospatial data. The agent is then presented with the column names, their data types, and the top 20 rows of the table.

\item Files with \textit{.xlsx} format: In this case, we use the \textit{pandas}\footnote{Available at: \url{https://pandas.pydata.org/}} library to handle this file format. For files containing multiple sheets, we provide the agent with all sheet names, the data types of columns in each sheet, and the top 20 rows from each sheet.

\item Files with \textit{.npz} format: In this case, we utilize the \textit{numpy}\footnote{Available at: \url{https://numpy.org/}} library to load the data. The agent is then presented with all keys and their corresponding values within this data structure.

\item Files with \textit{.cdf} format: In this case, we utilize the \textit{cdflib}\footnote{Available at: \url{https://cdflib.readthedocs.io/en/latest/}} library to load the data. For presentation, we call the \texttt{cdf\_info} and \texttt{globalattsget} functions on the loaded data structure, concatenate their outputs, and provide the result to the agent.

\item Any other data format: In this case, we open the files using Pythons \texttt{open} function and present the first 20 lines of the file to the agent.

\end{itemize}

\paragraph{Inference Setup.} We limit the maximum number of actions taken by the main agent to $T = 10$. For decoding, we use nucleus sampling \citep{Holtzman2020The} with a temperature of $\tau = 0.1$. Proprietary models are accessed through Vertex AI,\footnote{\url{https://cloud.google.com/vertex-ai?hl=en}} while open-source models are served using the vLLM library.\footnote{\url{https://docs.vllm.ai/en/latest/}} At each generation step, we cap the output length at 8,192 tokens. We evaluate three proprietary LLMs—Gemini 2.5 Pro, Gemini 2.5 Flash \citep{comanici2025gemini25pushingfrontier}, and Claude 4 Opus \citep{anthropic2025claude4}—alongside an open-source model specialized for code generation, Qwen3-Coder-30B-A3B-Instruct \citep{qwen3technicalreport}.\footnote{\url{https://huggingface.co/Qwen/Qwen3-Coder-30B-A3B-Instruct}} Experiments with open-source models are conducted on 2 NVIDIA A100 GPUs (80GB VRAM each) with 128GB RAM.

\newpage
\section{Examples and Case Studies}
\label{app:case-study}

This section presents several case studies highlighting different aspects of the Blackboard system.

\begin{figure}
    \centering
    \includegraphics[width=\textwidth]{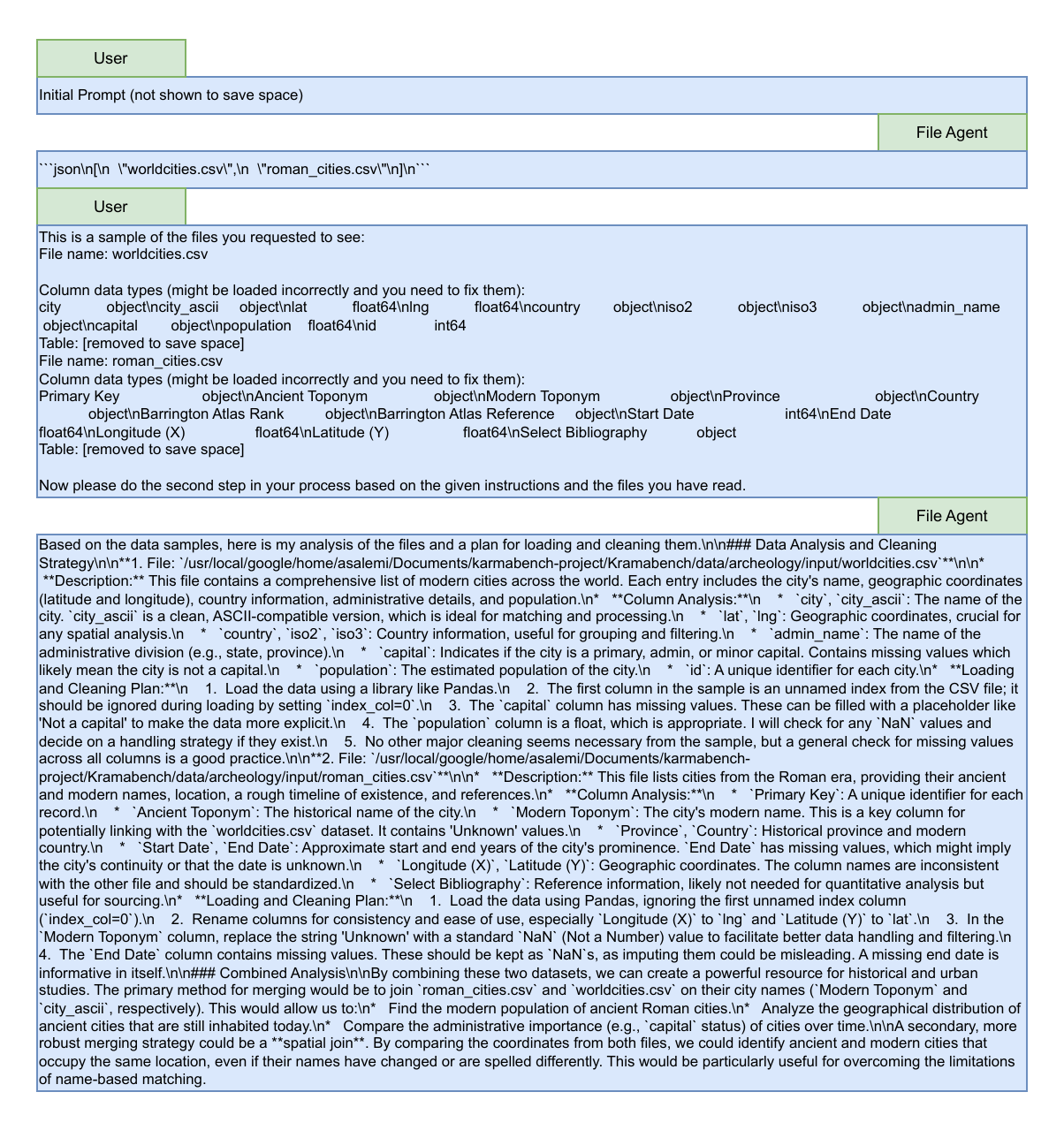}
    \vspace{-1.2cm}
    \caption{An example analyzing files by the file agent.}
    \label{fig:file-agent-analyze-example}
\end{figure}

\begin{figure}
    \centering
    \includegraphics[width=\textwidth]{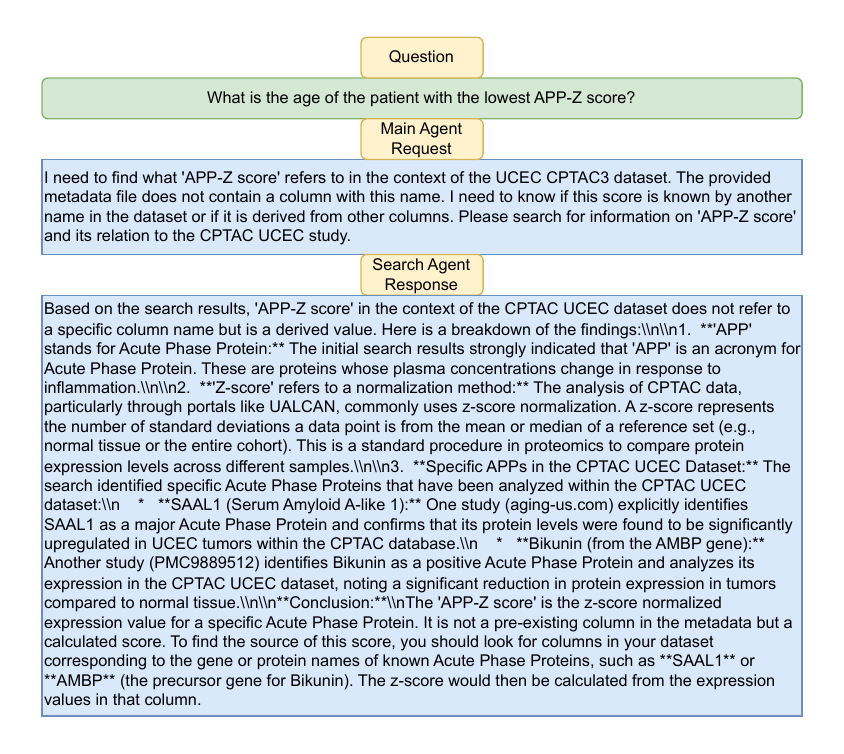}
    \vspace{-1.2cm}
    \caption{An example of the request by the main agent that the search agent has provided a guideline based on search results.}
    \label{fig:search-example-1}
\end{figure}

\begin{figure}
    \centering
    \includegraphics[width=\textwidth]{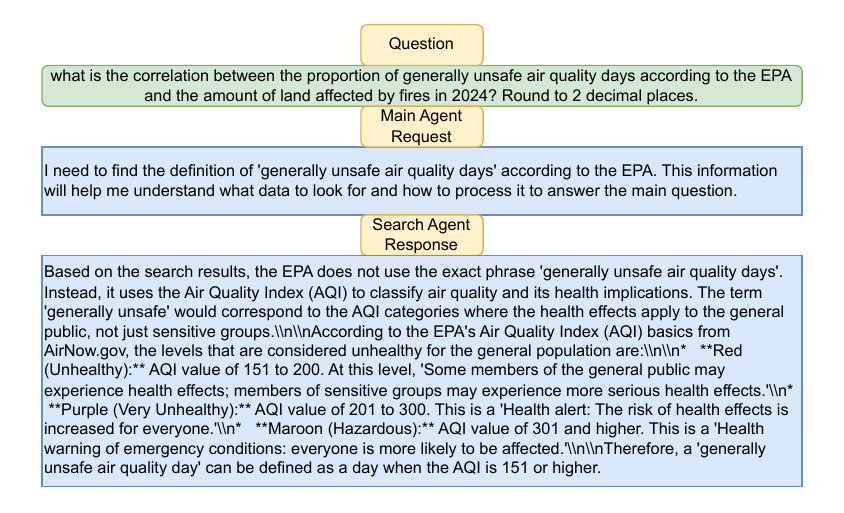}
    \vspace{-1.2cm}
    \caption{An example of the request by the main agent that the search agent has provided a guideline based on search results.}
    \label{fig:search-example-2}
\end{figure}

Figure~\ref{fig:file-agent-analyze-example} illustrates an example where a file agent requests access to files and performs their analysis. Figures~\ref{fig:search-example-1} and~\ref{fig:search-example-2} illustrate scenarios where the main agent lacked domain-specific knowledge and therefore posted requests on the blackboard seeking relevant information. In these cases, the search agent contributed by retrieving the necessary knowledge from the web, enabling the system to proceed with problem solving, which shows the effectiveness of search agent in problem solving.

\begin{figure}[h!]
    \centering
    \includegraphics[width=\textwidth]{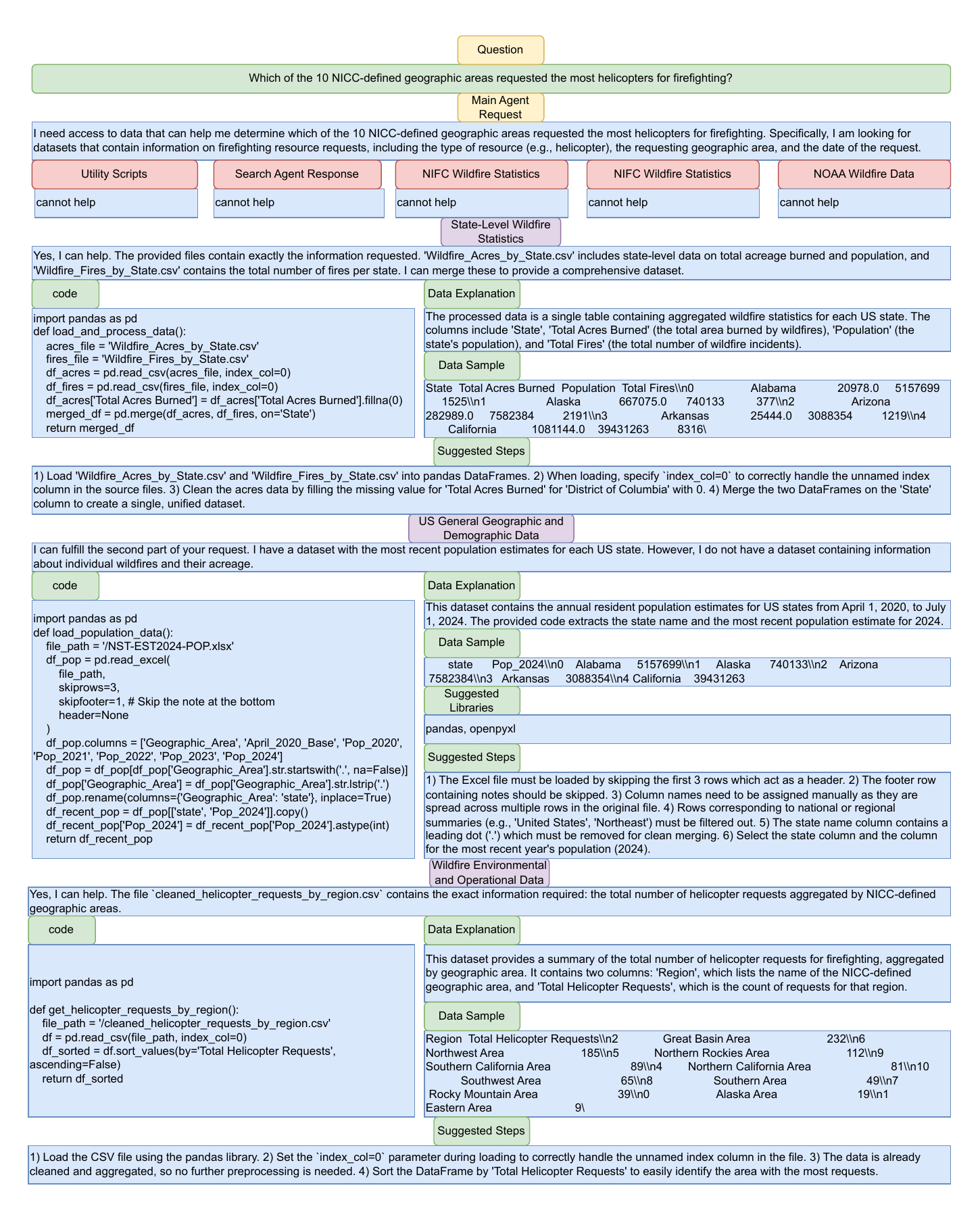}
    \vspace{-1cm}
    \caption{An example of the generated request by the blackboard system.}
    \label{fig:request-example}
\end{figure}

Another example of a blackboard request is shown in Figure~\ref{fig:request-example}. In this example, specifically file agents responded to the request. Here, the main agent, given a data science question, formulated a request specifying the likely column names and data formats required, along with guidance for interpretation. In response, three out of eight helper agents contributed. Although the relevant files were spread across different clusters managed by separate file agents, each responding agent independently provided file addresses, code snippets for loading the data, explanations of the structure, and suggested preprocessing steps. Together, these contributions encompassed all the ground-truth files needed to solve the problem. This case demonstrates how the main agent can effectively leverage the blackboard to coordinate decentralized knowledge and achieve accurate data discovery.

\begin{figure}[h!]
    \centering
    \includegraphics[width=\textwidth]{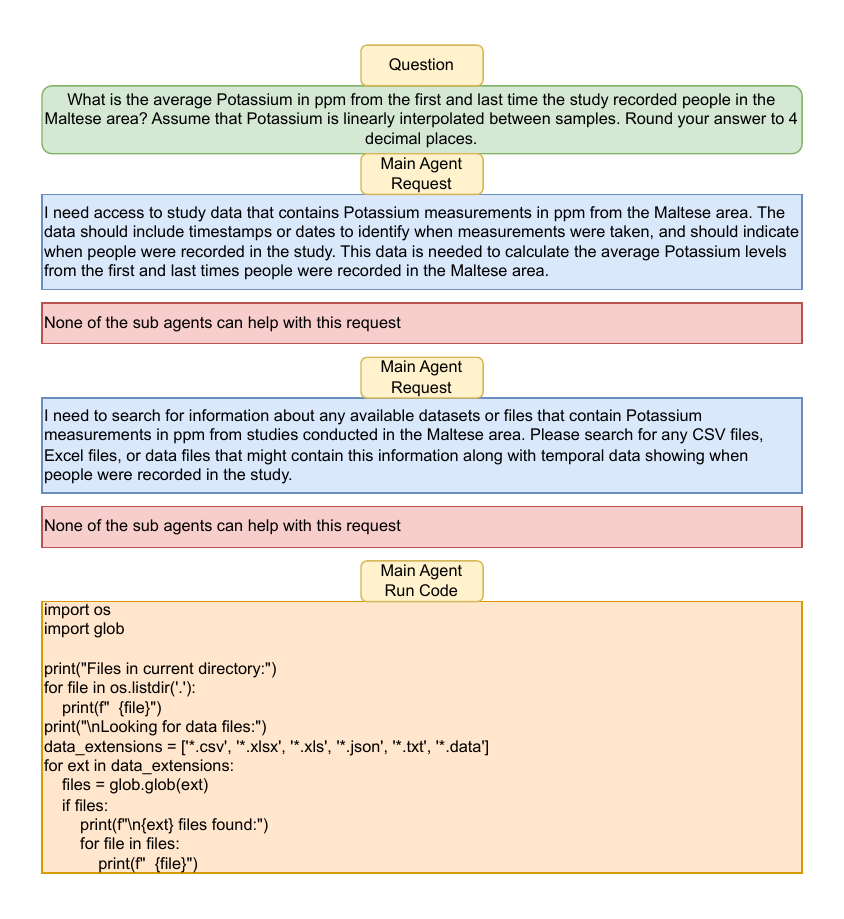}
    \vspace{-1cm}
    \caption{An example of the requesting operation that none of the sub-agents can help with this request. In this case, the main agent retries multiple times with different reformulations of the request, and upon repeated failure, resorts to alternative actions, such as executing code to locate the relevant files autonomously.}
    \label{fig:search-fail}
\end{figure}

In cases where none of the sub-agents can address a given request, as illustrated in Figure~\ref{fig:search-fail}, the main agent attempts multiple reformulations of the query to facilitate the retrieval of relevant sub-information or related content. If these attempts remain unsuccessful, the agent resorts to alternative strategies, such as searching local files through code generation. If it still fails to locate or infer the necessary information from the partially gathered evidence, it ultimately terminates the search.

\begin{figure}[h!]
    \centering
    \includegraphics[width=\textwidth]{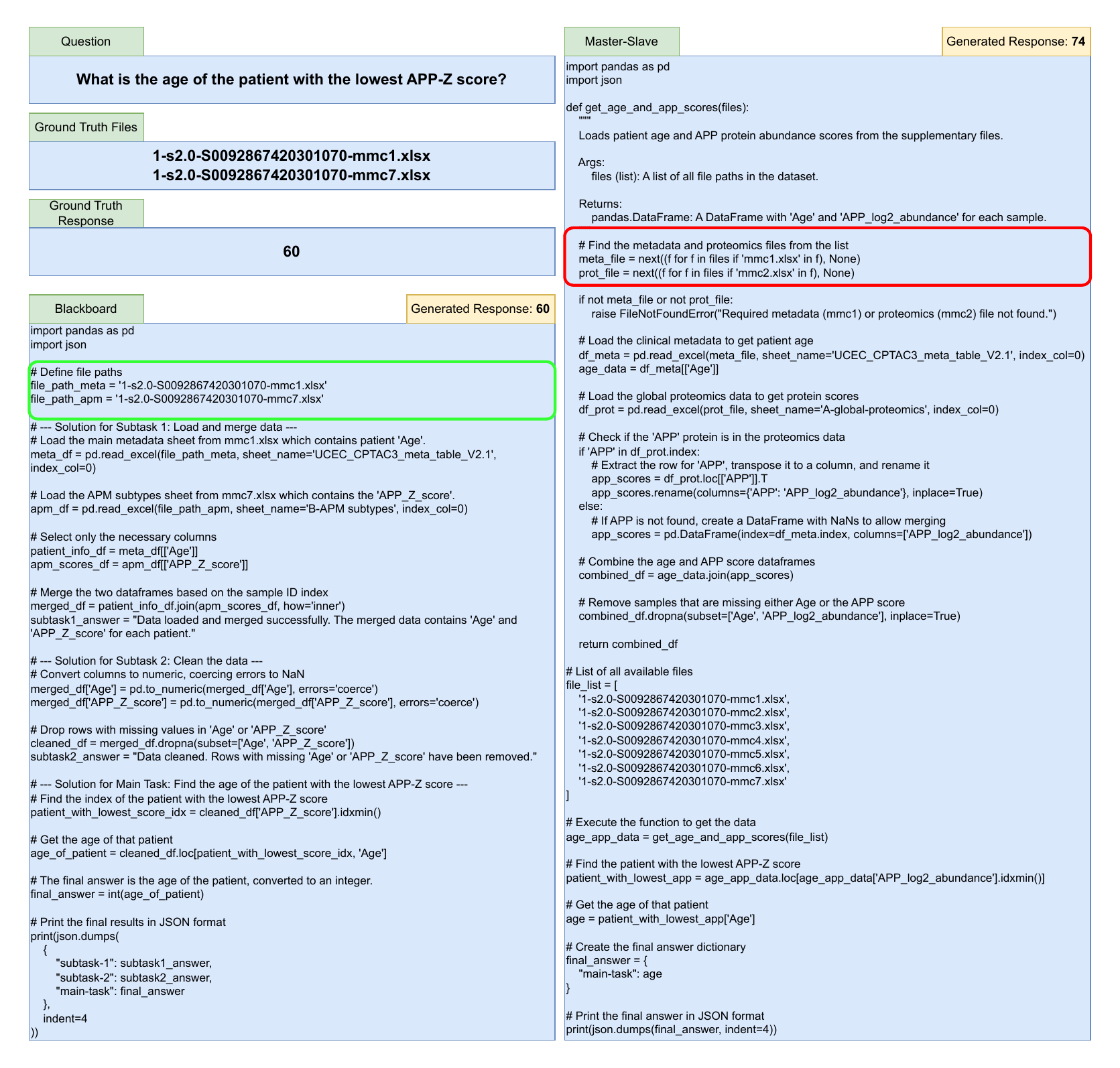}
    \vspace{-1cm}
    \caption{An example of the generated program by the blackboard system and master-slave system. The green box highlights where the blackboard system correctly selected the relevant files from the data lake, while the red box indicates where the master–slave system made an incorrect selection.}
    \label{fig:program-example}
\end{figure}

Figure~\ref{fig:program-example} compares programs generated by the Blackboard and Master–Slave systems. The Blackboard agent produced the correct solution by accurately interpreting the prompt and selecting the appropriate data files. Specifically, it identified that the patients \texttt{Age} was located in \texttt{mmc1.xlsx} and that the requested \texttt{APP-Z score} was in \texttt{mmc7.xlsx}. By contrast, the Master–Slave agent misinterpreted the request and instead used a general protein abundance score (\texttt{APP\_log2\_abundance}) from the wrong file, \texttt{mmc2.xlsx}. This misstep resulted in an incorrect answer of \texttt{74}, whereas the Blackboard agents precise data discovery and reasoning yielded the correct answer of \texttt{60}.

\newpage

\newpage

\newpage

\newpage

\newpage

\newpage

\newpage

\begin{figure}[h!]
    \centering
    \includegraphics[width=0.75\textwidth]{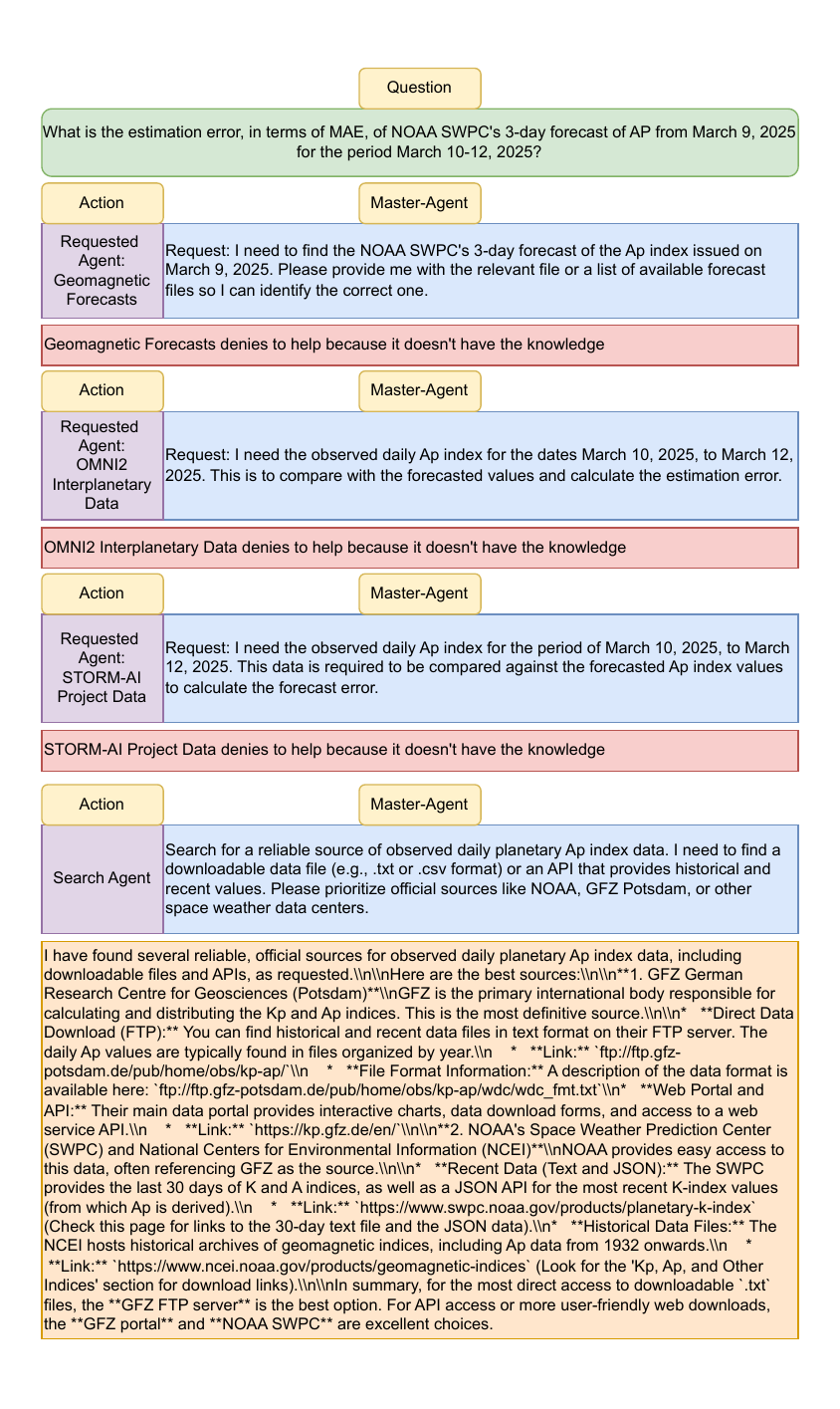}
    \vspace{-1cm}
    \caption{Example where the main controller in a Master–Slave system fails to route the query to the correct data sub-agent: after multiple retries, it falls back to the search agent, which returns only generic or incorrect information.}
    \label{fig:master-fail}
\end{figure}

\FloatBarrier

\newpage
\section{Further Results Analysis}
\label{app:results-analysis}

\begin{figure}
    \centering
    \includegraphics[width=0.9\textwidth]{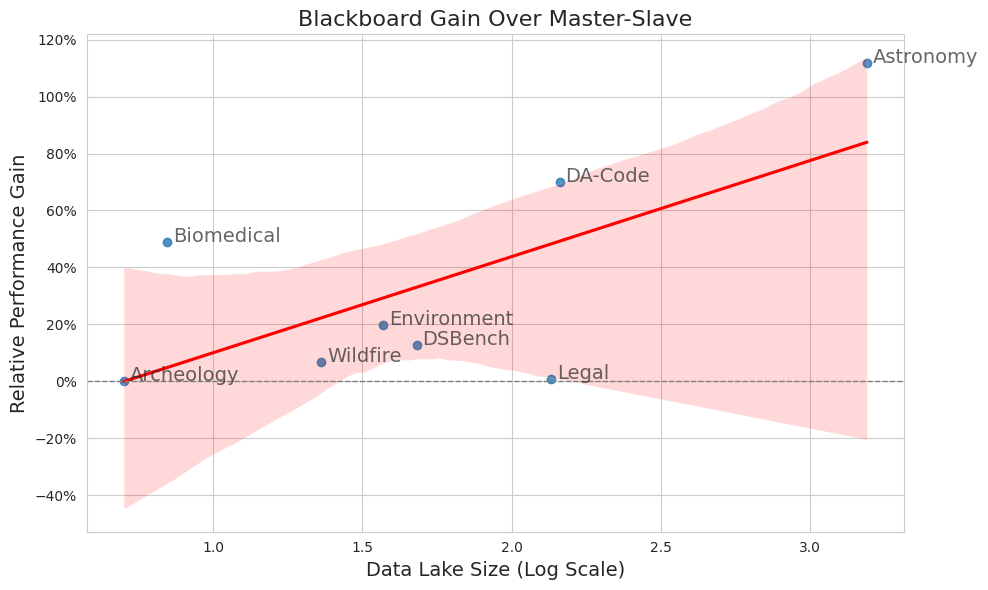}
    \caption{Relative performance gain of the Blackboard system over the Master–Slave system as a function of data lake size (log 10 scale). The red line represents the fitted regression line, with the shaded area showing the confidence interval. The results indicate a positive correlation between data lake size and the relative performance gain, suggesting that the Blackboard architecture scales more effectively with larger data lakes. This experiment is conducted using Gemini 2.5 Pro as the LLM.}
    \label{fig:performance-gain-datalake-size}
\end{figure}

\paragraph{Scalability Comparison between Blackboard and Master–Slave Systems:}
We report relative, rather than absolute, performance gains to ensure fair comparison across datasets with varying task difficulties and score ranges. Absolute improvements can be misleading when the baseline performance or achievable score range differs substantially---for instance, a 5-point gain might be minor in an easy task with high baseline scores but significant in a challenging one. Relative gain normalizes these differences by measuring proportional improvement with respect to the baseline, enabling consistent comparison across heterogeneous tasks. Using this normalized metric, we analyze the scalability of the Blackboard architecture compared to the Master–Slave baseline across datasets with different data lake sizes (Figure~\ref{fig:performance-gain-datalake-size}). Each point corresponds to a distinct task domain, and a fitted regression line reveals a clear positive correlation between data lake size and relative performance gain. This indicates that the Blackboard system scales more effectively as data environments grow larger, whereas the Master–Slave system exhibits limited scalability under such conditions.

\begin{figure*}[!t]
    \centering
    \includegraphics[width=0.85\textwidth]{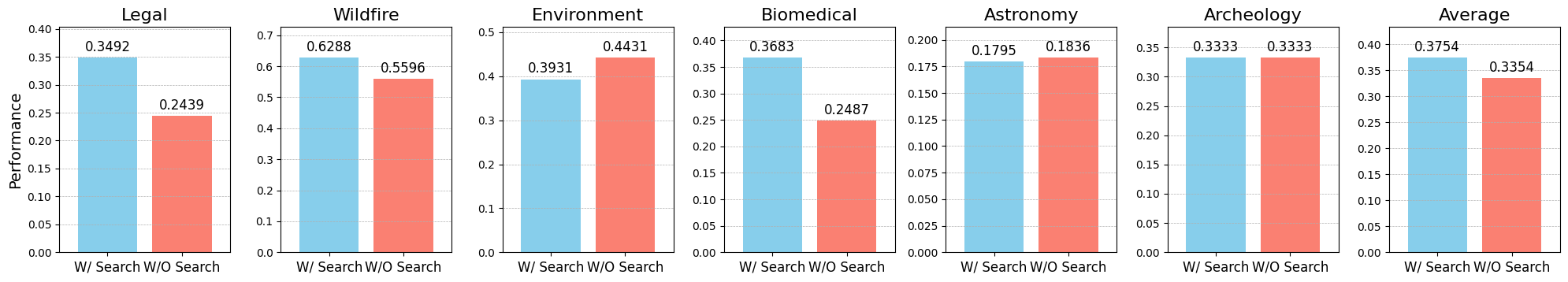}
    \vspace{-0.2cm}
    \caption{Performance of Blackboard System w/ and w/o search agent (Gemini 2.5 Pro).}
    \vspace{-0.4cm}
    \label{fig:search-wo-search}
\end{figure*}

\begin{figure*}[!t]
    \centering
    \includegraphics[width=0.85\textwidth]{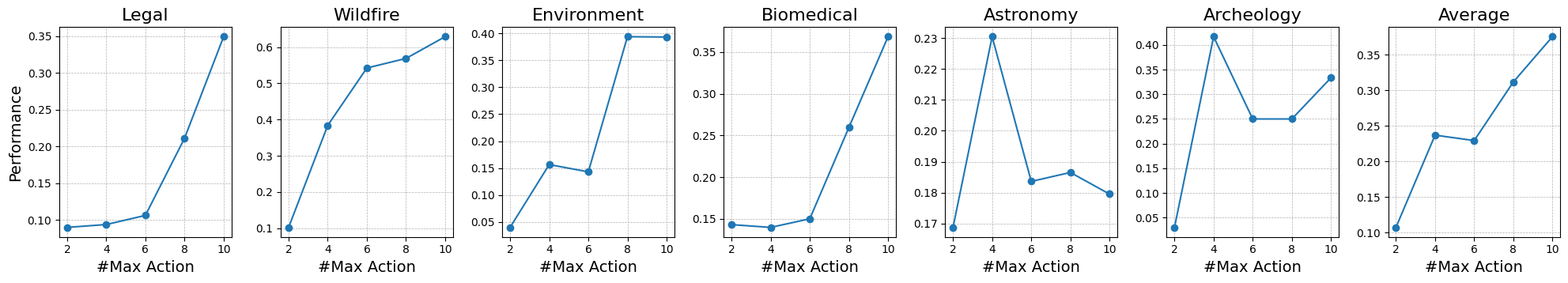}
    \vspace{-0.2cm}
    \caption{Performance of Blackboard System with various maximum actions by the main agent.}
    \label{fig:num-actions}
    \vspace{-0.5cm}
\end{figure*}

\begin{figure}
    \centering
    \includegraphics[width=0.9\textwidth]{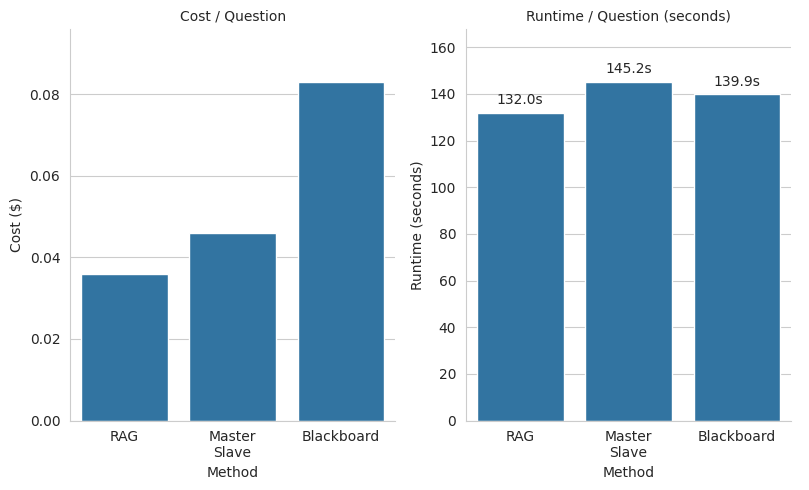}
    \vspace{-0.3cm}
    \caption{Runtime and cost analysis of the RAG, Master–Slave, and Blackboard systems on 50 examples from the KramaBench benchmark. We use Gemini 2.5 Pro as the LLM.}
    \label{fig:runtime-cost}
\end{figure}

\paragraph{Runtime and Cost Analysis:}

To assess the efficiency–cost trade-off of the Blackboard system relative to the RAG and Master–Slave baselines, we randomly sampled 50 questions from the KramaBench benchmark spanning all domains and measured both runtime and cost per question for each method. Unlike the RAG and Master–Slave baselines, which execute their component calls sequentially following the ReAct framework, the Blackboard architecture parallelizes sub-agent interactions: once the main agent posts a request to the shared blackboard, the corresponding sub-agents process it independently. As shown in Figure~\ref{fig:runtime-cost}, the runtime of all three systems lies in a narrow band (132.0--145.2 seconds), indicating no significant difference in latency. In terms of monetary cost, the Blackboard system is more expensive per question (approximately $2.3\times$ the cost of RAG and $1.8\times$ that of Master–Slave), reflecting its increased token usage. However, this additional cost translates into substantial performance gains---54.1\% over RAG and 18.8\% over Master–Slave---so that Blackboard delivers markedly better answer quality while maintaining comparable runtime, offering a favorable accuracy–cost trade-off.

\paragraph{Compare with Other More Advanced Data Science Baselines:}

To further evaluate our method against advanced data-science-oriented agent frameworks, we compare the Blackboard system with Data Interpreter \citep{hong-etal-2025-data} and AutoGen \citep{wu2023autogenenablingnextgenllm}, both designed for multi-agent data analysis and reasoning. The results on the KramaBench benchmark are presented in Table~\ref{tab:krama-gemini-pro}. As shown, the Blackboard system outperforms these baselines in five out of six tasks as well as on the overall average, highlighting its stronger adaptability and coordination capabilities across diverse data-science scenarios.

\begin{table}[t]
    \centering
    \caption{Results on the KramaBench benchmark using Gemini 2.5 Pro as the underlying LLM.}
    \adjustbox{max width=\textwidth}{
    \begin{tabular}{l|cccccc|c}
        \toprule
        Method & Archaeology & Astronomy & Biomedical & Environment & Legal & Wildfire & Average \\
        \midrule
        Data Interpreter & \textbf{41.67}\% & 12.72\% & 28.05\% & 9.87\% & 30.04\% & 59.67\% & 30.34\% \\
        AutoGen  & 16.67\% & 4.39\% & 7.25\% & 19.38\% & 26.38\% & 41.76\% & 19.30\% \\
        \midrule
        Blackboard    & {33.33}\% & \textbf{17.95}\% & \textbf{36.83}\% & \textbf{39.31}\% & \textbf{34.92}\% & \textbf{62.88}\% & \textbf{37.53}\% \\
        \bottomrule
    \end{tabular}}
    \label{tab:krama-gemini-pro}
\end{table}

\paragraph{Effect of Clustering Based on File Name and Content and Number of Clusters on the Performance:}

As described in \textsection\ref{sec:method}, we employ Gemini 2.5 Pro to perform file clustering based on filenames. However, given that file contents can be lengthy, it is impractical to provide the full text of each file to the model for clustering. To address this, we additionally experiment with content-based clustering using a semantic embedding model. Specifically, we encode each file’s content using E5-Large \citep{wang2022text} and apply the KMeans algorithm \citep{Lloyd1982LeastSQ} to partition the files into multiple clusters. Unlike Gemini, which automatically determines the number of clusters, KMeans requires this value to be specified. We evaluate settings with 2, 4, 8, and 16 clusters and use datasets containing at least 100 files—namely, the Legal and Astronomy subsets from the KramaBench and DA-Code datasets.

\begin{figure}
    \centering
    \includegraphics[width=\textwidth]{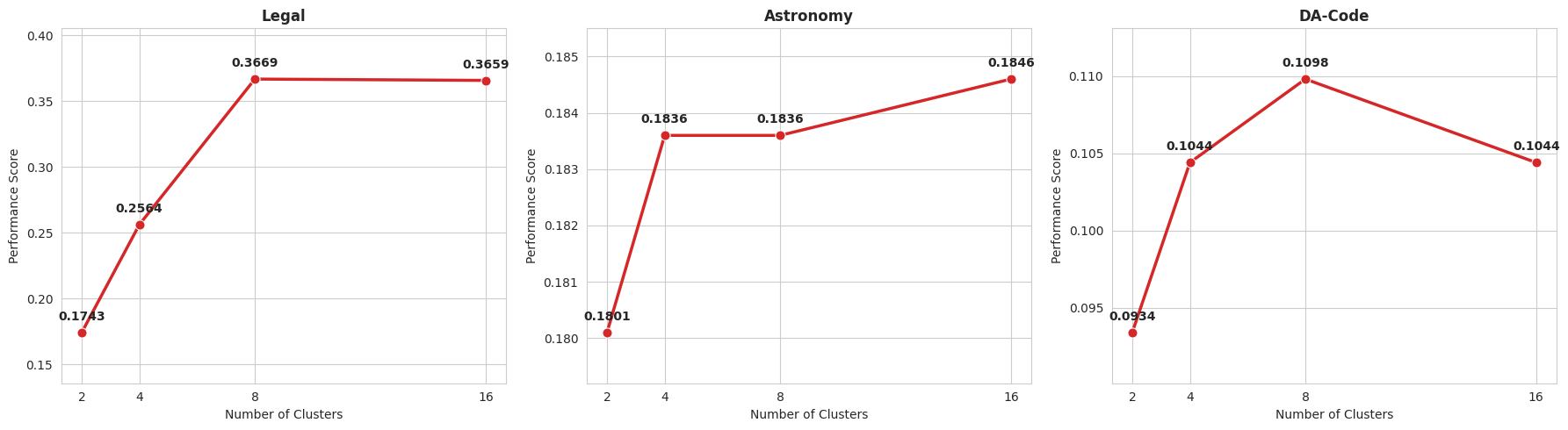}
    \caption{Effect of the number of clusters on the performance of the Blackboard system. Clustering is performed using the KMeans algorithm with E5-Large as the embedding model. We use Gemini 2.5 Pro as the LLM.}
    \label{fig:cluster-num}
\end{figure}

\begin{figure}
    \centering
    \includegraphics[width=\textwidth]{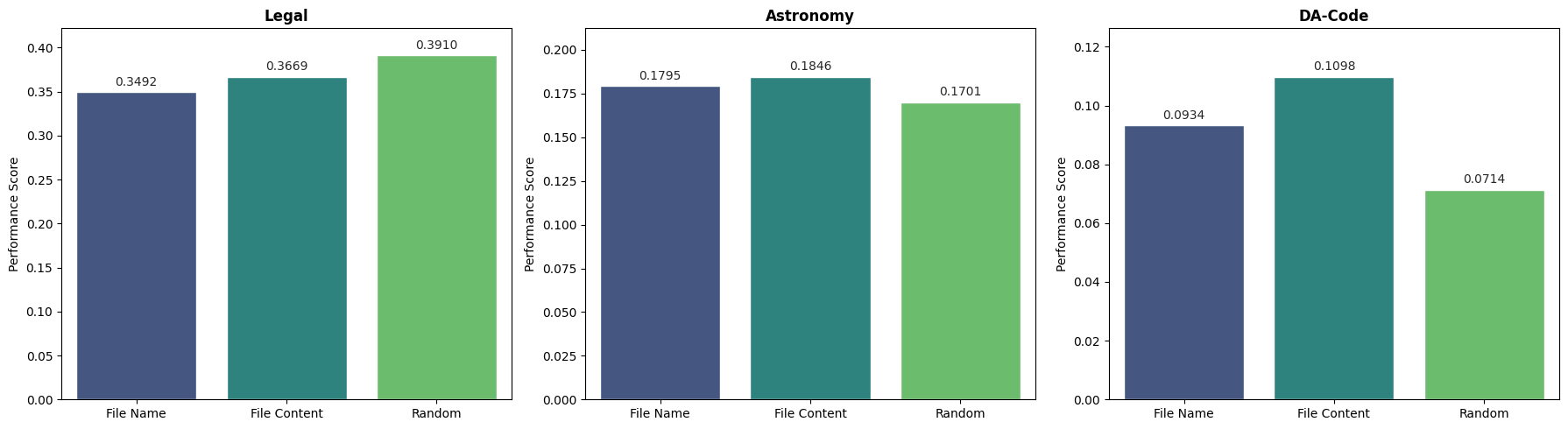}
    \caption{Effect of clustering based on filename using Gemini 2.5 Pro and based on file content using the KMeans algorithm with E5-Large as the embedding model. We use Gemini 2.5 Pro as the LLM.}
    \label{fig:file-vs-content}
\end{figure}

The results of the content-based clustering are presented in Figure~\ref{fig:cluster-num}. Overall, increasing the number of clusters leads to higher performance, as it allows each sub-agent to specialize on a smaller subset of files and perform more targeted reasoning. Furthermore, to compare filename-based clustering with Gemini 2.5 Pro against content-based clustering using E5-Large embeddings and random clustering, we report the results in Figure~\ref{fig:file-vs-content}. As shown, clustering based on semantic content yields consistently better performance than clustering by filename alone or random clustering. This indicates that our proposed framework generalizes effectively to embedding-based clustering and does not rely solely on large language models for grouping data files.

\newpage
\section{Example IDs for Filtered Datasets}
\label{app:dataset-ids}

We retained the following example IDs from the instances in the DSBench dataset:

\begin{multicols}{3}
    \begin{itemize}[nosep, leftmargin=*]
        \item \texttt{00000001\_question17}
        \item \texttt{00000001\_question7}
        \item \texttt{00000001\_question16}
        \item \texttt{00000001\_question13}
        \item \texttt{00000001\_question12}
        \item \texttt{00000001\_question9}
        \item \texttt{00000001\_question10}
        \item \texttt{00000001\_question11}
        \item \texttt{00000001\_question18}
        \item \texttt{00000001\_question15}

        \item \texttt{00000004\_question50}
        \item \texttt{00000004\_question45}
        \item \texttt{00000004\_question42}
        \item \texttt{00000004\_question48}
        \item \texttt{00000004\_question47}
        \item \texttt{00000004\_question49}
        \item \texttt{00000004\_question44}
        \item \texttt{00000004\_question41}
        \item \texttt{00000004\_question43}

        \item \texttt{00000005\_question21}
        \item \texttt{00000005\_question29}
        \item \texttt{00000005\_question28}
        \item \texttt{00000005\_question24}
        \item \texttt{00000005\_question25}
        \item \texttt{00000005\_question23}
        \item \texttt{00000005\_question26}
        \item \texttt{00000005\_question20}
        \item \texttt{00000005\_question27}

        \item \texttt{00000006\_question17}
        \item \texttt{00000006\_question22}
        \item \texttt{00000006\_question21}
        \item \texttt{00000006\_question18}
        \item \texttt{00000006\_question24}
        \item \texttt{00000006\_question25}
        \item \texttt{00000006\_question19}
        \item \texttt{00000006\_question23}
        \item \texttt{00000006\_question26}
        \item \texttt{00000006\_question20}

        \item \texttt{00000007\_question17}
        \item \texttt{00000007\_question22}
        \item \texttt{00000007\_question7}
        \item \texttt{00000007\_question16}
        \item \texttt{00000007\_question6}
        \item \texttt{00000007\_question5}
        \item \texttt{00000007\_question13}
        \item \texttt{00000007\_question12}
        \item \texttt{00000007\_question21}
        \item \texttt{00000007\_question9}
        \item \texttt{00000007\_question10}
        \item \texttt{00000007\_question11}
        \item \texttt{00000007\_question3}
        \item \texttt{00000007\_question18}
        \item \texttt{00000007\_question4}
        \item \texttt{00000007\_question19}
        \item \texttt{00000007\_question23}
        \item \texttt{00000007\_question14}
        \item \texttt{00000007\_question2}
        \item \texttt{00000007\_question15}
        \item \texttt{00000007\_question20}
        \item \texttt{00000007\_question1}
        \item \texttt{00000007\_question8}

        \item \texttt{00000008\_question33}
        \item \texttt{00000008\_question31}
        \item \texttt{00000008\_question28}
        \item \texttt{00000008\_question32}
        \item \texttt{00000008\_question34}

        \item \texttt{00000010\_question17}
        \item \texttt{00000010\_question7}
        \item \texttt{00000010\_question16}
        \item \texttt{00000010\_question6}
        \item \texttt{00000010\_question5}
        \item \texttt{00000010\_question13}
        \item \texttt{00000010\_question12}
        \item \texttt{00000010\_question9}
        \item \texttt{00000010\_question10}
        \item \texttt{00000010\_question11}
        \item \texttt{00000010\_question3}
        \item \texttt{00000010\_question18}
        \item \texttt{00000010\_question4}
        \item \texttt{00000010\_question19}
        \item \texttt{00000010\_question14}
        \item \texttt{00000010\_question2}
        \item \texttt{00000010\_question15}
        \item \texttt{00000010\_question20}
        \item \texttt{00000010\_question1}
        \item \texttt{00000010\_question8}

        \item \texttt{00000011\_question7}
        \item \texttt{00000011\_question6}
        \item \texttt{00000011\_question5}
        \item \texttt{00000011\_question3}
        \item \texttt{00000011\_question4}
        \item \texttt{00000011\_question2}
        \item \texttt{00000011\_question1}
        \item \texttt{00000011\_question8}

        \item \texttt{00000012\_question7}
        \item \texttt{00000012\_question6}
        \item \texttt{00000012\_question5}
        \item \texttt{00000012\_question9}
        \item \texttt{00000012\_question10}
        \item \texttt{00000012\_question3}
        \item \texttt{00000012\_question4}
        \item \texttt{00000012\_question2}
        \item \texttt{00000012\_question1}
        \item \texttt{00000012\_question8}

        \item \texttt{00000013\_question17}
        \item \texttt{00000013\_question22}
        \item \texttt{00000013\_question16}
        \item \texttt{00000013\_question21}
        \item \texttt{00000013\_question18}
        \item \texttt{00000013\_question24}
        \item \texttt{00000013\_question19}
        \item \texttt{00000013\_question23}
        \item \texttt{00000013\_question20}

        \item \texttt{00000016\_question7}
        \item \texttt{00000016\_question6}
        \item \texttt{00000016\_question5}
        \item \texttt{00000016\_question12}
        \item \texttt{00000016\_question9}
        \item \texttt{00000016\_question11}
        \item \texttt{00000016\_question3}
        \item \texttt{00000016\_question2}
        \item \texttt{00000016\_question1}
        \item \texttt{00000016\_question8}

        \item \texttt{00000017\_question39}
        \item \texttt{00000017\_question33}
        \item \texttt{00000017\_question38}
        \item \texttt{00000017\_question29}
        \item \texttt{00000017\_question37}
        \item \texttt{00000017\_question30}
        \item \texttt{00000017\_question32}
        \item \texttt{00000017\_question35}
        \item \texttt{00000017\_question34}
        \item \texttt{00000017\_question36}

        \item \texttt{00000018\_question24}
        \item \texttt{00000018\_question25}
        \item \texttt{00000018\_question23}

        \item \texttt{00000019\_question6}
        \item \texttt{00000019\_question13}
        \item \texttt{00000019\_question9}
        \item \texttt{00000019\_question10}
        \item \texttt{00000019\_question14}
        \item \texttt{00000019\_question15}

        \item \texttt{00000020\_question33}
        \item \texttt{00000020\_question31}
        \item \texttt{00000020\_question29}
        \item \texttt{00000020\_question37}
        \item \texttt{00000020\_question30}
        \item \texttt{00000020\_question28}
        \item \texttt{00000020\_question32}
        \item \texttt{00000020\_question35}
        \item \texttt{00000020\_question34}
        \item \texttt{00000020\_question36}

        \item \texttt{00000022\_question39}
        \item \texttt{00000022\_question33}
        \item \texttt{00000022\_question38}
        \item \texttt{00000022\_question40}
        \item \texttt{00000022\_question37}
        \item \texttt{00000022\_question35}
        \item \texttt{00000022\_question34}
        \item \texttt{00000022\_question36}

        \item \texttt{00000025\_question39}
        \item \texttt{00000025\_question33}
        \item \texttt{00000025\_question38}
        \item \texttt{00000025\_question40}
        \item \texttt{00000025\_question37}
        \item \texttt{00000025\_question35}
        \item \texttt{00000025\_question41}
        \item \texttt{00000025\_question34}
        \item \texttt{00000025\_question36}

        \item \texttt{00000027\_question13}
        \item \texttt{00000027\_question12}
        \item \texttt{00000027\_question14}
        \item \texttt{00000027\_question15}

        \item \texttt{00000029\_question7}
        \item \texttt{00000029\_question6}
        \item \texttt{00000029\_question5}
        \item \texttt{00000029\_question9}
        \item \texttt{00000029\_question10}
        \item \texttt{00000029\_question3}
        \item \texttt{00000029\_question4}
        \item \texttt{00000029\_question2}
        \item \texttt{00000029\_question1}
        \item \texttt{00000029\_question8}

        \item \texttt{00000030\_question7}
        \item \texttt{00000030\_question6}
        \item \texttt{00000030\_question5}
        \item \texttt{00000030\_question3}
        \item \texttt{00000030\_question4}
        \item \texttt{00000030\_question2}
        \item \texttt{00000030\_question1}

        \item \texttt{00000032\_question7}
        \item \texttt{00000032\_question6}
        \item \texttt{00000032\_question5}
        \item \texttt{00000032\_question3}
        \item \texttt{00000032\_question4}
        \item \texttt{00000032\_question2}
        \item \texttt{00000032\_question1}

        \item \texttt{00000033\_question7}
        \item \texttt{00000033\_question6}
        \item \texttt{00000033\_question3}
        \item \texttt{00000033\_question4}
        \item \texttt{00000033\_question2}
        \item \texttt{00000033\_question1}
        \item \texttt{00000033\_question8}

        \item \texttt{00000034\_question7}
        \item \texttt{00000034\_question16}
        \item \texttt{00000034\_question6}
        \item \texttt{00000034\_question5}
        \item \texttt{00000034\_question13}
        \item \texttt{00000034\_question12}
        \item \texttt{00000034\_question9}
        \item \texttt{00000034\_question10}
        \item \texttt{00000034\_question11}
        \item \texttt{00000034\_question3}
        \item \texttt{00000034\_question4}
        \item \texttt{00000034\_question14}
        \item \texttt{00000034\_question2}
        \item \texttt{00000034\_question15}
        \item \texttt{00000034\_question1}
        \item \texttt{00000034\_question8}

        \item \texttt{00000035\_question7}
        \item \texttt{00000035\_question6}
        \item \texttt{00000035\_question5}
        \item \texttt{00000035\_question9}
        \item \texttt{00000035\_question3}
        \item \texttt{00000035\_question4}
        \item \texttt{00000035\_question2}
        \item \texttt{00000035\_question1}
        \item \texttt{00000035\_question8}

        \item \texttt{00000038\_question5}
        \item \texttt{00000038\_question3}
        \item \texttt{00000038\_question2}
        \item \texttt{00000038\_question1}

        \item \texttt{00000043\_question17}
        \item \texttt{00000043\_question7}
        \item \texttt{00000043\_question16}
        \item \texttt{00000043\_question6}
        \item \texttt{00000043\_question5}
        \item \texttt{00000043\_question13}
        \item \texttt{00000043\_question12}
        \item \texttt{00000043\_question9}
        \item \texttt{00000043\_question10}
        \item \texttt{00000043\_question11}
        \item \texttt{00000043\_question3}
        \item \texttt{00000043\_question18}
        \item \texttt{00000043\_question4}
        \item \texttt{00000043\_question19}
        \item \texttt{00000043\_question14}
        \item \texttt{00000043\_question2}
        \item \texttt{00000043\_question15}
        \item \texttt{00000043\_question20}
        \item \texttt{00000043\_question1}
        \item \texttt{00000043\_question8}
    \end{itemize}
\end{multicols}

We retained the following example IDs from the instances in the DA-Code dataset:

\begin{multicols}{3}
    \begin{itemize}[nosep, leftmargin=*]
        \item \texttt{di-text-001}
        \item \texttt{di-text-002}
        \item \texttt{di-text-003}
        \item \texttt{di-text-004}
        \item \texttt{di-text-005}
        \item \texttt{di-text-006}
        \item \texttt{di-text-007}
        \item \texttt{di-text-008}
        \item \texttt{di-text-009}
        \item \texttt{di-text-010}
        \item \texttt{di-text-011}
        \item \texttt{di-text-012}
        \item \texttt{di-text-013}
        \item \texttt{di-text-014}
        \item \texttt{di-text-015}
        \item \texttt{di-text-016}
        \item \texttt{di-text-017}
        \item \texttt{di-text-018}
        \item \texttt{di-text-019}
        \item \texttt{di-text-020}
        \item \texttt{di-text-021}
        \item \texttt{di-text-023}
        \item \texttt{di-text-024}
        \item \texttt{di-text-025}
        \item \texttt{di-text-027}
        \item \texttt{di-text-028}
        \item \texttt{di-text-029}
        \item \texttt{di-text-030}
        \item \texttt{di-text-031}
        \item \texttt{di-text-032}
        \item \texttt{di-text-033}
        \item \texttt{di-text-034}
        \item \texttt{di-text-035}
        \item \texttt{di-text-036}
        \item \texttt{di-text-037}
        \item \texttt{di-text-038}
        \item \texttt{di-text-039}
        \item \texttt{di-text-040}
        \item \texttt{di-text-041}
        \item \texttt{di-csv-001}
        \item \texttt{di-csv-002}
        \item \texttt{di-csv-003}
        \item \texttt{di-csv-005}
        \item \texttt{di-csv-006}
        \item \texttt{di-csv-007}
        \item \texttt{di-csv-008}
        \item \texttt{di-csv-009}
        \item \texttt{di-csv-010}
        \item \texttt{di-csv-011}
        \item \texttt{di-csv-012}
        \item \texttt{di-csv-013}
        \item \texttt{di-csv-014}
        \item \texttt{di-csv-015}
        \item \texttt{di-csv-016}
        \item \texttt{di-csv-017}
        \item \texttt{di-csv-018}
        \item \texttt{di-csv-019}
        \item \texttt{di-csv-020}
        \item \texttt{di-csv-021}
        \item \texttt{di-csv-022}
        \item \texttt{di-csv-023}
        \item \texttt{dm-csv-001}
        \item \texttt{dm-csv-002}
        \item \texttt{dm-csv-009}
        \item \texttt{dm-csv-010}
        \item \texttt{dm-csv-011}
        \item \texttt{dm-csv-016}
        \item \texttt{dm-csv-019}
        \item \texttt{dm-csv-020}
        \item \texttt{dm-csv-021}
        \item \texttt{dm-csv-028}
        \item \texttt{dm-csv-029}
        \item \texttt{dm-csv-032}
        \item \texttt{dm-csv-034}
        \item \texttt{dm-csv-038}
        \item \texttt{dm-csv-041}
        \item \texttt{dm-csv-049}
        \item \texttt{dm-csv-050}
        \item \texttt{dm-csv-055}
        \item \texttt{data-sa-011}
        \item \texttt{data-sa-012}
        \item \texttt{data-sa-022}
        \item \texttt{data-sa-034}
        \item \texttt{data-sa-035}
        \item \texttt{data-sa-037}
        \item \texttt{data-sa-041}
        \item \texttt{data-sa-044}
        \item \texttt{data-sa-047}
        \item \texttt{data-sa-065}
        \item \texttt{data-sa-067}
        \item \texttt{data-sa-068}
    \end{itemize}
\end{multicols}

\newpage
\section{Large Language Model Usage for Writing}

In this paper, we employ LLMs---specifically Gemini and ChatGPT---as general-purpose writing tools. Draft text is provided to these models, which are then asked to improve the writing by correcting grammatical errors and refining the structure. The edited text is then verified and edited if needed. The use of LLMs in this paper is limited strictly to text refinement. They were not employed for tasks such as generating any new content or references.

\end{document}